\documentclass{article}
\usepackage{graphicx} 
\usepackage[a4paper,left=2.73cm,right=2.7cm,top=3cm,bottom=3.5cm]{geometry}

\usepackage{amsfonts,amsmath,amssymb,tabstackengine}
\usepackage[normalem]{ulem}
\usepackage{tikz}
\usepackage{array}
\allowdisplaybreaks
\usepackage{slashed}
\global\long\def\not#1{\slashed{#1}}%
\usepackage{tensor}

\usepackage{comment}

\usepackage{cite}

\definecolor{Gray}{gray}{0.95}

\usepackage[colorlinks=true,linktocpage=true,linkcolor=blue,citecolor=blue]{hyperref}
\usepackage{graphicx}

\addtocontents{toc}{\protect\setcounter{tocdepth}{2}}
\numberwithin{equation}{section}
\usepackage{upgreek}
\usepackage{mathrsfs}
\usepackage{dsfont}
\global\long\def\ri{\mathrm{i}}%
\global\long\def\dd{\mathrm{d}}
\global\long\def\Vol{\mathrm{Vol}}
\global\long\def\bP{\mathds{P}}

\global\long\def\bR{\mathbb{R}}%
\global\long\def\bZ{\mathbb{Z}}%

\global\long\def\hM{\mathscr{M}}%

 \usepackage [latin1]{inputenc}
 
\begin{document}

\begin{titlepage}
	\thispagestyle{empty}
	\begin{center}
	    { \LARGE{\bf  Supersymmetric Warped Solutions\\ \vspace{0.4cm} from Type IIB Orientifold Reduction }}
		
		\vspace{50pt}
		
		{S.~Maurelli $^{1,2}$, R.~Noris$^{3}$, M.~Oyarzo$^{1,4}$, H.~Samtleben$^{5,6}$ and M.~Trigiante$^{1,2}$}
		
		\vspace{25pt}

		{
		$^1${\it  Department of Applied Science and Technology, Politecnico di Torino, Corso Duca degli Abruzzi, 24, 10129 Torino, Italy}

		\vspace{15pt}

		$^2${\it  INFN, Sezione di Torino, Via P. Giuria 1, 10125 Torino, Italy}

		\vspace{15pt}

		$^3${\it CEICO, Institute of Physics of the Czech Academy of Sciences,
Na Slovance 2, 182 21 Prague 8, Czech Republic.}

		\vspace{15pt}

		$^4${\it  Departamento de F\'isica, Universidad de Concepci\'on Casilla, 160-C, Concepci\'on, Chile}
		}
		
		\vspace{15pt}
		
		$^5${\it ENSL, CNRS, Laboratoire de Physique, F-69342 Lyon, France
}
		
		\vspace{15pt}
		
		$^6${\it
Institut Universitaire de France (IUF), France}
		
		\vspace{40pt}
		
		{ABSTRACT}
	\end{center}
    We construct a family of supersymmetric solutions in Type IIB supergravity of the form ${\rm WAdS}_3\times {\rm WS}^3\times T^4$, where ${\rm WAdS}_3$  and ${\rm WS^3}$ denote a warped anti-de Sitter spacetime and a warped 3-sphere, respectively, while $T^4$ denotes an internal 4-torus. These backgrounds are constructed by uplifting corresponding solutions in the $D=6$, $\mathcal{N}=(1,1)$ ungauged supergravity describing the closed string zero-modes of Type IIB supergravity compactified on a $T^4/\mathbb{Z}_2$-orientifold. More specifically, the supersymmetric solutions are ${\rm WAdS}_3\times {\rm WS}^3\times T^4$ with lightlike warped AdS$_3$ and ${\rm WAdS}_3\times {\rm S}^3\times T^4$ in which the warping of AdS$_3$ is generic. Moreover, we also construct solutions in the form of a warped product $\mathrm{LM}^3_{\zeta,\omega}\times_{{\rm w}} \mathrm{S}^3\times T^4$ of a 2-parameter deformation $\mathrm{LM}^3_{\zeta,\omega}$ of ${\rm AdS}_3$ and a three-sphere. We discuss the relation of these backgrounds to known solutions.
\end{titlepage}
\date{January 2025}

\tableofcontents

\section{Introduction}

String and supergravity backgrounds containing AdS$_3 \times \rm S^3$ factors, arising as the near-horizon geometry of D1-D5 brane configurations in type IIB string theory, have provided key examples of holographic dualities since the early days of the field~\cite{Maldacena:1998bw,Deger:1998nm,deBoer:1998kjm}. In certain cases, they have even paved the way for a 
full-fledged derivation of the holographic duality~\cite{Eberhardt:2019ywk}.
A natural class of deformations of these setups involves squashing the $\rm S^3$ sphere factor as well as warping the AdS$_3$ geometry. 
Geometrically, squashing of the sphere corresponds to breaking the full ${\rm SO}(4) \simeq {\rm SU}(2)_L \times {\rm SU}(2)_R$ isometry of the round three-sphere $\rm S^3$ down to a subgroup, typically ${\rm SU}(2)_L \times {\rm U}(1)_R$, or even further. If the AdS$_3$ factor remains undeformed, the dual theory is still a two-dimensional conformal field theory (CFT) --- albeit one with reduced global symmetry. The holographic dictionary remains intact, and observables such as correlation functions, central charges, and entropy can be computed within the usual AdS/CFT framework.

By contrast, deforming the {AdS}$_3$ factor itself requires a modification of the standard holographic framework. Warped AdS$_3$ (WAdS$_3$) geometries break the full ${\rm SO}(2,2)$ isometry group of {AdS}$_3$ down to ${\rm SL}(2,\mathbb{R}) \times {\rm U}(1)$, and can be viewed as non-trivial fibrations over AdS$_2$, while breaking the original Lorentz invariance. Their holographic duals rather correspond to deformations of CFT, so-called dipole or warped CFTs \cite{Anninos:2008fx,Guica:2010sw,Song:2011sr,Detournay:2012pc,Afshar:2015wjm}. Such theories, while lacking full conformal symmetry, retain enough structure to permit a consistent holographic dictionary, including Cardy-like entropy formulas and modular properties.

Much of the development of this warped AdS/CFT correspondence has taken place in the context of three-dimensional topologically massive gravity (TMG). Crucially, both squashed and stretched AdS$_3 \times \rm S^3$ geometries also arise as fully consistent string theory backgrounds \cite{Israel:2004vv,Detournay:2005fz,Azeyanagi:2012zd}. In type IIB supergravity, such solutions can be embedded explicitly by turning on appropriate fluxes. They have been constructed explicitly directly in ten dimensions or within consistent truncations to six and to three dimensions \cite{Bobev:2011qx,El-Showk:2011euy,Eloy:2023acy,Eloy:2024lwn,Deger:2024xnd,Deger:2024obg}. In general, these solutions only preserve part of the original supersymmetries.\par
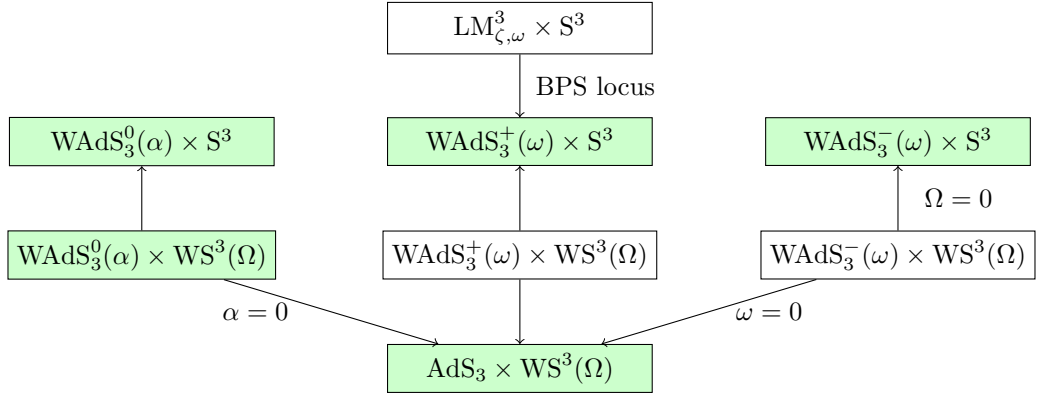
\begin{figure}
    \centering
\begin{tikzpicture}[align=center,minimum height=0.5cm,minimum width=3.5cm]
\node[rectangle,draw] (t0) at (0,0) {$\mathrm{LM}^3_{\zeta,\omega} \times \rm S^3$};
\node[rectangle,draw, fill=green!20] (c2) at (0,-1.5) {$\mathrm{WAdS}_{3}^{+}(\omega)\times\mathrm{S}^{3}$};
\node[rectangle,draw, fill=green!20] (c3) at (5,-1.5) {$\mathrm{WAdS}_{3}^{-}(\omega)\times\mathrm{S}^{3}$};
\node[rectangle,draw, fill=green!20] (c1) at (-5,-1.5) {$\mathrm{WAdS}_{3}^{0}(\alpha)\times\mathrm{S}^{3}$};
\node[rectangle,draw, fill=green!20] (b1) at (-5,-3) {$\mathrm{WAdS}_{3}^{0}(\alpha)\times\mathrm{\rm{WS}}^{3}(\Omega)$};
\node[rectangle,draw] (b2) at (0,-3) {$\mathrm{WAdS}_{3}^{+}(\omega)\times\mathrm{\rm{WS}}^{3}(\Omega)$};
\node[rectangle,draw] (b3) at (5,-3) {$\mathrm{WAdS}_{3}^{-}(\omega)\times\mathrm{\rm{WS}}^{3}(\Omega)$};
\node[rectangle,draw,fill=green!20] (h2) at (0,-4.5) {$\mathrm{AdS}_{3}\times\mathrm{\rm{WS}}^{3}(\Omega)$};
\draw[solid,->] (t0) -- (c2)
  node[midway,xshift=10mm] {BPS locus};
\draw[solid,->] (b1) -- (c1) node[midway,yshift=3mm] {};
\draw[solid,->] (b2) -- (c2)
  node[midway,yshift=3mm] {};
  \draw[solid,->] (b3) -- (c3)
  node[midway,xshift=8mm] {$\Omega=0$};
    \draw[solid,->] (b3) -- (h2)
  node[midway,xshift=8mm] {$\omega=0$};
    \draw[solid,->] (b2) -- (h2);
    \draw[solid,->] (b1) -- (h2)node[midway,xshift=-10mm] {$\alpha=0$};
\end{tikzpicture}
    \caption{Representation of the solutions discussed in this work and of their relations. Green boxes correspond to supersymmetric solutions.}
    \label{fig:enter-label}
\end{figure}

One of the main results of this work is the construction of a class of solutions to Type IIB supergravity, of the form ${\rm WAdS}_3\times {\rm WS}^3\times T^4$. They are graphically represented in Figure \ref{fig:enter-label}. All these solutions can be described within an effective, ungauged $D=6$ $\mathcal{N}=(1,1)$ theory describing the closed string sector of the zero-modes of  Type IIB supergravity on a $T^4/\mathbb{Z}_2$-orientifold (see \cite{Angelantonj:2002ct} for a general review). Being this model a consistent truncation of the maximal $D=6$ $\mathcal{N}=(2,2)$, our six-dimensional solutions uplift to backgrounds of the form ${\rm WAdS}_3\times {\rm WS}^3\times T^4$. The backgrounds with the smallest residual symmetry ${\rm SL}(2,\mathbb{R})\times {\rm SU}(2)\times {\rm U}(1)^2$ are the double-warped ones, of which only those with lightlike warping are supersymmetric and preserve 4 supercharges (i.e. $1/8$ of the $D=10$ supersymmetries). 
As far as the generic double-warped solution is concerned, the warping parameters of the two three-dimensional subspaces are independent; this allows us to consider limits in which such parameters are separately set to zero, and the corresponding spaces undeformed. As illustrated in Figure \ref{fig:enter-label}, in the round 3-sphere limit of the double-warped solutions, supersymmetry is restored, independently of the kind of warping in ${\rm WAdS}_3 $. We also construct a background (at the top of the figure) of the form of a warped product $\mathrm{LM}^3_{\zeta,\omega}\times_{{\rm w}} \mathrm{S}^3$ of a 2-parameter deformation $\mathrm{LM}^3_{\zeta,\omega}$ of ${\rm AdS}_3$ and a three-sphere. This background is the ``mirror'' of the non-supersymmetric 2-parameter background constructed in \cite{Eloy:2021fhc}, in which the deformation is on the anti-de Sitter side. Just as for their counterparts in \cite{Eloy:2021fhc}, restricting the two parameters to a suitable locus, supersymmetry is restored. We provide the explicit ten-dimensional expression of the double-warped solutions. \par As a bonus, we give the detailed embedding of the $D=6$ $\mathcal{N}=(1,1)$ closed string sector of the orientifold theory in Type IIB supergravity as well as the supersymmetry variations of its fermion fields. These relations allow for a straightforward computation of the explicit form of all six-dimensional backgrounds given here as solutions to the ten-dimensional theory. \par
\par
Instances of supersymmetric, double-warped solutions can be found in the literature \cite{El-Showk:2011euy,Hoare:2022asa}, in the compactification of Type IIB superstring theory on $K3$ and $T^4$, respectively. In our double-warped solutions, the independent deformations of the three-dimensional subspaces are related to the non-vanishing of the components of the RR 4-form and the Kalb-Ramond 2-form along odd cycles of $T^4$. Thus, we argue that analogous configurations cannot be obtained in a $K3$ compactification, since this manifold features no odd-cycles. As for the double-warped supersymmetric solutions in \cite{Hoare:2022asa}, they were obtained through a TsT transformation of the ${\rm AdS}_3\times {\rm S}^3$ background and feature the same deformation parameter on the two three-dimensional subspaces. Such configurations only contain self-dual 3-form tensor field strengths and no vectors. As a consequence of this, they cannot be obtained as limits of the double-warped solutions considered here, since the latter also contain vector fields and, even in the limit of equal warping parameters of the two three-dimensional subspaces, where vector fields vanish, feature 3-form field strengths with no definite self-duality property. \par

The paper is organised as follows. In Section 2, we define the six-dimensional $\mathcal{N}=(1,1)$ theory describing the closed string sector of the reduction of Type IIB supergravity on a $T^4/\mathbb{Z}_2$-orientifold. In Section 3, we provide, for the reader's convenience, a unified review of the relevant properties of the warped spaces ${\rm WAdS}_3$ and ${\rm WS}^3$, and establish the conventions.  Readers familiar with these notions can skip to Sections 4, 5, and 6, where the new solutions are discussed. We end with some concluding remarks. Appendix \ref{appA} is devoted to the definition of the relevant conventions, and in Appendix \ref{appB}, the main facts about Type IIB supergravity and its dimensional reduction on the $T^4/\mathbb{Z}_2$-orientifold are reviewed, restricting our attention to the closed string sector only.

\section{The six-dimensional theory}\label{section2}
We will consider here the half-maximal  $\mathcal N=(1,1)$, $D=6$ ungauged supergravity, originating as a compactification of the Type IIB, $D=10$ model on a $T^{4}/ \mathbb{Z}_{2}$-orientifold, i.e. in the presence of $O5/D5$ sources. We shall not consider the open string modes but restrict ourselves to the closed string sector only. The field content of the theory is described by a gravity multiplet coupled to four vector multiplets, defining the following coset geometry for the scalar manifold
\begin{equation}\label{ScalarManifold}
    \mathcal{M}_{scal} = \frac{G}{H} = {\rm O}(1,1) \times \frac{{\rm O}(4,4)}{{\rm O}(4)\times {\rm O}(4)}.
\end{equation}
\\
The bosonic action reads 
\begin{align}\label{bosonicaction}
I_{\mathrm{6D}} & =\frac{1}{2\kappa_{6}^{2}}\int\Big[\mathcal{R}\star1-\frac{1}{2}e^{\phi}\star F_{(3)}\wedge F_{(3)}\nonumber\\
&-\frac{1}{2}e^{-\frac{1}{2}\phi}(\hat{\gamma}^{ab}\star H_{(2)a}\wedge H_{(2)b}+\hat{\gamma}_{ab}\star\check{\mathcal{F}}^{a}_{(2)}\wedge\check{\mathcal{F}}^{b}_{(2)})+\dd\check{\mathcal{C}}^{a}_{(2)}\wedge H_{(2)a}\wedge C_{(2)}\nonumber \\
&-\frac{1}{4}\star\dd\phi\wedge\dd\phi+\frac{1}{4}\star\dd\hat{\gamma}_{ab}\wedge\dd\hat{\gamma}^{ab}-\frac{1}{4}\hat{\gamma}\hat{\gamma}^{ac}\hat{\gamma}^{bd}\star\dd C_{(1)ab}\wedge\dd C_{(1)cd}\Big]\,,
\end{align}
where the indices $a,b,\dots=1,\dots, 4$, label the coordinates $y^a$ of the internal 4-torus, see Appendix \ref{appB}.
Here the ${\rm O}(1,1)$ factor is parametrized by the dilaton $\phi$, while the coset $\frac{{\rm O}(4,4)}{{\rm O}(4)\times {\rm O}(4)}$ is spanned by the moduli $\hat{\gamma}_{ab}=\hat{\gamma}_{ba},\,C_{ab}=-C_{ba}$, coming from the internal metric  $\upgamma_{ab}$ and from the reduction of the ten-dimensional RR 2-form potential, respectively. The eight vector fields $(\check{\mathcal{C}}^{a}_{(1)},\, B_{(1)a})$, originate from the toroidal reduction of the RR 4-form field and the Kalb-Ramond fields, respectively, and transform in the fundamental representation of ${\rm O}(4,4)$. Only the ${\rm O}(1,1)$ factor acts non-trivially on the 3-form field strength and its dual. We refer the reader to Appendix \ref{appB} for further details on this truncation and for the precise relation between the six-dimensional fields and the ten-dimensional ones. In particular, the relation between the field $\phi$ and the ten-dimensional dilaton $\Phi$, as well as the expression of $\upgamma_{ab}$ in terms of the matrix $\hat{\gamma}_{ab}$ entering the above Lagrangian, are given in eq.s \eqref{B.20}. 
\\
The bosonic equations of motion descending from the action \eqref{bosonicaction} are
\begin{align}\label{eqmotion1}
    \delta_\phi S=0:\qquad & 0=
    \dd\star \dd\phi-\frac12 e^{-\frac{\phi}{2}}\left(\hat\gamma^{ab}\star H_{(2)a}\wedge H_{(2)b}+\hat\gamma_{ab}\star\check{\mathcal{F}}^{a}_{(2)}\wedge \check{\mathcal{F}}^{b}_{(2)}\right)+e^{\phi}\star F_{(3)}\wedge F_{(3)}\nonumber\\
    \delta_{C_{(2)}} S=0:\qquad & 0=\dd(e^\phi \star F_{(3)})-\dd\check{\mathcal C}^a_{(2)}\wedge H_{(2)a}\nonumber\\
    \delta_{C_{ab}} S=0:\qquad & 0=\dd\left(\hat\gamma\hat\gamma^{ac}\hat\gamma^{bd}\star \dd C_{cd}\right)-\frac12\hat\gamma_{pq}\star\check{\mathcal F}^{p}_{(2)}\wedge H_{(2)c}\epsilon^{abcq}\nonumber\\
    \delta_{B_{(1)a}} S=0:\qquad & 0=\dd\left(e^{-\frac{\phi}{2}}\hat\gamma^{ab}\star H_{(2)a}-\frac12e^{-\frac{\phi}{2}}\hat\gamma_{ac}\star\check{\mathcal{F}}^{a}_{(2)}C_{de}\epsilon^{bcde}-\dd \check{\mathcal C}^b_{(2)}\wedge C_{(2)}\right)\nonumber\\
    \delta_{\check{\mathcal{C}}_{(1)}^a} S=0:\qquad & 0=\dd(e^{-\frac{\phi}{2}}\hat\gamma_{ab}\star\check{\mathcal{F}}^a_{(2)}-H_{(2)b}\wedge C_{(2)})\nonumber\\
    \delta_{\gamma_{ab}} S=0:\qquad &
    0= e^{-\frac{\phi}{2}}(\star H_{(2)a}\wedge H_{(2)b}-\hat\gamma_{ac}\hat \gamma_{bd}\star\check{\mathcal{F}}^{c}_{(2)}\wedge\check{\mathcal{F}}^{d}_{(2)})-\frac12\dd\star\dd \hat\gamma_{ab}+\frac12\dd\star\dd\hat\gamma^{cd}\hat\gamma_{ca}\hat\gamma_{db}\nonumber\\
    &-\frac12\hat\gamma\hat\gamma_{ab}\hat \gamma^{ce}\hat\gamma^{df}\star\dd C_{cd}\wedge \dd C_{ef}+\hat\gamma\hat\gamma^{cd}\star\dd C_{ac}\wedge \dd C_{bd}
\end{align}
together with the equation for the metric, which reads
\begin{align}\label{eqmotion2}
    &R_{\mu\nu}-\frac14 e^\phi F_{\mu\rho\sigma} F_{\nu}{}^{\rho\sigma}-\frac12e^{-\frac{\phi}{2}}(\hat\gamma^{ab}H_{a\mu\rho}H_{b\nu}{}^\rho+\hat\gamma_{ab}\check{\mathcal{F}}^{a}_{\mu\rho}\check{\mathcal{F}}^{b}_{\nu}{}^{\rho})-\frac14\partial_\mu \phi\partial_\nu\phi+\frac14\partial_\mu\hat \gamma_{ab}\partial_\nu\hat \gamma^{ab}\nonumber\\
    &-\frac14\hat\gamma\hat\gamma^{ac}\hat\gamma^{bd}\partial_\mu C_{ab}\partial_\nu C_{cd}+\frac14 g_{\mu\nu}\bigg( \frac16e^\phi F_{\rho\sigma\lambda}F^{\rho\sigma\lambda}+\frac14e^{-\frac{\phi}{2}}(\hat\gamma^{ab}H_{a\rho\sigma}H_b^{\rho\sigma}+\hat\gamma_{ab}\check{\mathcal{F}}^{a}_{\rho\sigma}\check{\mathcal{F}}^{b\rho\sigma})\bigg) =0\,.
\end{align}
\paragraph{${\rm O}(4,4)$-Covariant notation}
The scalar fields $\upvarphi=(\upvarphi^s)\equiv \{\hat{\gamma}_{ab},\,C_{ab}\}$ are moduli of the undeformed ${\rm AdS}_3\times {\rm S}^3$ background describing the near-horizon geometry of the D1-D5 system. As mentioned above, they span the coset manifold ${\rm O}(4,4)/[{\rm O}(4)\times {\rm O}(4)]$. The six-dimensional action and field equations can be rewritten in a manifestly ${\rm O}(4,4)$-covariant way, by encoding their dependence on the scalars $\upvarphi^s$ in the following ${\rm O}(4,4)$-coset representative:
$$\mathcal{V}_{\cal M}{}^{\underline{\cal N}}=\left(\begin{matrix}{\bf 1} & {\bf 0}\cr \frac{1}{2}\epsilon^{abcd}\,C_{cd} & {\bf 1}\end{matrix}\right)\cdot \left(\begin{matrix}\hat{E}_a{}^{\underline{a}} & {\bf 0}\cr {\bf 0} & (\hat{E}^{-T}){}^a{}_{\underline{a}}\end{matrix}\right)\,\,,\,\,\,\,(\mathcal{M},\mathcal{N},\ldots=1,\dots 8)\,,$$
where 
$$\hat{\gamma}_{ab}=\sum_{\underline{a}=1}^4\hat{E}_a{}^{\underline{a}}\hat{E}_b{}^{\underline{a}}\,.$$ 
In this basis of the fundamental representation of ${\rm O}(4,4)$, the invariant matrix, and its inverse, read:
\begin{equation}
\mathcal{I}_{\cal MN}=\mathcal{I}^{\cal MN}=\left(\begin{matrix}{\bf 0} & {\bf 1} \cr {\bf 1} & {\bf 0}\end{matrix}\right)\,.
\end{equation}
It is convenient to introduce the following symmetric, positive definite, pseudo-orthogonal matrix, function of the point on the coset:
\begin{equation}
    \mathbb{M}(\upvarphi)\equiv \mathcal{V}(\upvarphi)\cdot \mathcal{V}(\upvarphi)^T\in {\rm O}(4,4)\,.
\end{equation}
Grouping the vector fields in an ${\rm SO}(4,4)$-vector:
$$\mathbb{B}^{\cal M}\equiv (\check{\mathcal{C}}^a,\,B_a)\,,$$
the kinetic terms for the scalars $\upvarphi^{s}$ and for the vector fields can be recast in the following ${\rm O}(4,4)$-invariant form:
$$-\frac{1}{2}\star \dd \upvarphi^s\wedge \dd\upvarphi^r\,\mathscr{G}_{rs}(\upvarphi)-\frac{e^{-\frac{\phi}{2}}}{2}\star \mathbb{F}_{(2)}^T\cdot\mathbb{M}(\upvarphi)\wedge \mathbb{F}_{(2)}=\frac{1}{8} {\rm Tr}\left(\star\dd \mathbb{M}\wedge \dd \mathbb{M}^{-1}\right)-\frac{e^{-\frac{\phi}{2}}}{2}\star \mathbb{F}_{(2)}^T\cdot\mathbb{M}(\upvarphi)\wedge\mathbb{F}_{(2)}\,, $$
where we have suppressed the pseudo-orthogonal indices $\mathcal{M,\,N},\dots$, defined $\mathbb{F}_{(2)}\equiv (\mathbb{F}_{(2)}^\mathcal{M})=(\dd \mathbb{B}^\mathcal{M})$ and denoted by $\mathscr{G}_{rs}(\upvarphi)$ the Riemannian metric on ${\rm O}(4,4)/[{\rm O}(4)\times {\rm O}(4)]$.\par
The field equations for the scalar fields $\upvarphi^s$ read:
\begin{equation}
    \nabla_\mu\left(\partial^\mu\upvarphi^s\right)+\tilde{\Gamma}(\upvarphi)_{s_1 s_2}^s\,\partial_\mu\upvarphi^{s_1}\partial^\mu\upvarphi^{s_2}=\frac{e^{-\frac{\phi}{2}}}{4}\,\mathscr{G}^{sr}(\upvarphi)\,\mathbb{F}_{\mu\nu}^T\cdot\frac{\partial}{\partial\upvarphi^r}\mathbb{M}(\upvarphi)\cdot\mathbb{F}^{\mu\nu}\,,\label{eqgamma}
\end{equation}
where  $\tilde{\Gamma}(\upvarphi)$ denotes the Levi-Civita symbol associated with $\mathscr{G}_{rs}$.
In the explicit solutions, which will be given in the sequel, the scalar fields will be chosen to be constant and fixed at the origin $O$ of the coset manifold 
$$\upvarphi\equiv O\,\,\,\Leftrightarrow\,\,\,\,\hat{\gamma}_{ab}=\delta_{ab}\,\,,\,\,\,C_{ab}=0\,.$$
The consistency of this position requires the source term on the right-hand side of \eqref{eqgamma} to vanish on the background:
$$\mathbb{F}_{\mu\nu}^T\cdot\left.\frac{\partial}{\partial\upvarphi^r}\mathbb{M}\right\vert_{\upvarphi=O}\cdot\mathbb{F}^{\mu\nu}=0\,.$$
This is guaranteed by the following condition on $\mathbb{F}_{\mu\nu}$:
$$\mathbb{F}_{(2)}=\mathcal{I}\cdot \mathbb{F}_{(2)}\,\,\Rightarrow\,\,\,\,\dd \check{\mathcal{C}}_{(1)}^a=\dd B_{(1)\,a}\,,$$
which holds on the solutions.\par
Using the ${\rm O}(4,4)$-covariance of the field equations, we can construct backgrounds with generic constant values $\upvarphi_0$ of $\upvarphi$ by transforming the fields of the original solution by the corresponding coset-representative $\mathcal{V}(\upvarphi_0)$:
$$\mathbb{F}_{(2)}\,\longrightarrow\,\,\,\mathbb{F}_{(2)}'=(\mathcal{V}(\upvarphi_0)^{-1})^T\cdot \mathbb{F}_{(2)}\,.$$
The transformed vector field strengths now satisfy the condition:
$$\mathbb{F}_{(2)}'=\mathbb{M}(\upvarphi_0)^{-1}\cdot \mathcal{I}\cdot \mathbb{F}_{(2)}'\,,$$
which still implies the vanishing of the source term in the equation for $\upvarphi^s$:
$$\mathbb{F}_{\mu\nu}^{\prime T}\cdot\left.\frac{\partial}{\partial\upvarphi^r}\mathbb{M}\right\vert_{\upvarphi=\upvarphi_0}\cdot\mathbb{F}^{\prime\,\mu\nu}=\mathbb{F}_{\mu\nu}^{\prime T}\cdot\left.\frac{\partial}{\partial\upvarphi^r}\mathbb{M}\cdot \mathbb{M}^{-1}\right\vert_{\upvarphi=\upvarphi_0}\cdot \mathcal{I}\cdot\mathbb{F}^{\prime\,\mu\nu}=0\,,$$
where we have used the property of the matrix $\frac{\partial}{\partial\upvarphi^r}\mathbb{M}\cdot \mathbb{M}^{-1}\cdot \mathcal{I}$ of being antisymmetric.

\subsection{The Killing spinor equations}
Let us discuss here the supersymmetry variations of the fermionic fields on a generic bosonic background, that is, for vanishing fermionic fields. We give here the supersymmetry variations of the ten-dimensional fermionic fields $\boldsymbol{\lambda}$ and $\boldsymbol{\Psi}_M$, written in terms of the dimensionally reduced fields. The conventions on Dirac matrices and spinors are illustrated in Appendices \ref{appA} and \ref{appB}. In particular, the transformation for the spin-$\frac12$ field reads 
\begin{align}\label{susylambda}
    \delta\boldsymbol{\lambda}&=\frac{1}{2}e^{\frac{\phi}{8}}\hat{\gamma}^{-\frac{1}{16}}\,\mathds 1_8\otimes \Gamma_{*}\otimes \sigma_3\left(\frac{1}{2}\not{\dd}\phi\otimes\mathds 1_4\otimes\sigma_{3}+\frac{1}{2}\not{\dd}\log\sqrt{\hat{\gamma}}\otimes\mathds 1_4\otimes\sigma_{3} \right.\nonumber\\
    &\left.- \frac{1}{2}e^{-\frac{\phi}{4}}\not H_{(2)a}\otimes\hat{E}_{\underline{a}}{}^{a}\Gamma_{*}\Gamma^{\underline{a}}\otimes \mathds 1_2-\frac{1}{2}e^{\frac{\phi}{2}}\not F_{(3)}\otimes\mathds 1_4\otimes\ri\sigma_{2}-\frac{1}{4}\hat{\gamma}^{\frac12}\not{\dd}C_{ab}\otimes\hat{E}_{\underline{a}}{}^{a}\hat{E}_{\underline{b}}{}^{b}\Gamma^{\underline{ab}}\otimes\ri\sigma_{2}\right)\boldsymbol{\epsilon}\,,
\end{align}
where $\hat{E}_{\underline{a}}{}^{a}$ are the inverse vielbein matrices of the metric $\hat \gamma_{ab}$. The supersymmetry transformation of the four spin-$\frac12$ fields coming from the reduction of the ten-dimensional gravitino are given by
\begin{align}\label{susypsia}
    \delta\boldsymbol{\Psi}_a&=\bigg[\frac{\hat\gamma^{\frac18}}{4}\not{\dd}(\hat\gamma^{-\frac38}\hat\gamma_{ab}e^{\frac{\phi}{4}})\otimes \Gamma^{b}\Gamma_* \otimes \mathds 1_2+\frac{1}{8}\hat\gamma^{-\frac14}\check{\not{\mathcal{F}}}{}^{\underline b}_{(2)}\otimes\Gamma_{\underline b}\Gamma_{a}\Gamma_*\otimes\ri \sigma_{2}\nonumber\\
    &-\frac{1}{16}e^{\frac{3\phi}{4}}\hat\gamma^{-\frac{1}{4}}\not F_{(3)}\otimes\Gamma_{a}\Gamma_*\otimes\sigma_1+\frac{1}{16}\hat\gamma^{-\frac14}(\not H_{(2)\underline b}\otimes\Gamma_{a}\Gamma^{\underline b}-4\not H_{(2)a}\otimes\mathds{1}_{4})\otimes\sigma_{3}\nonumber\\
    &-\frac{1}{16}e^{\frac\phi4}\hat\gamma^{\frac14}(\not F_{\underline{bc}}\otimes\Gamma_a\Gamma^{\underline{bc}}\Gamma_*-4\not F_{a\underline{b}}\otimes\Gamma^{\underline b}\Gamma_*)\otimes \sigma_1\bigg]\boldsymbol{\epsilon}\,.
\end{align}
Finally, the supersymmetry variation for gravitino can be written as
\begin{align}\label{susypsimu}
    \delta\boldsymbol{\Psi}_{\mu}\dd x^{\mu} & =\dd \boldsymbol{\epsilon}+\mathbf{W}\boldsymbol{\epsilon}\,,
\end{align}
with 
\begin{align}
    \mathbf{W}&=\frac{1}{4}\gamma^{\underline{\rho\sigma}}(\omega_{\underline{\rho\sigma}}+2e_{\underline{\rho}}\partial_{\underline\sigma}\log\Delta)\otimes\mathds{1}_{8}+\frac14\mathds{1}_{8}\otimes\Gamma^{\underline{ab}}(\hat E_{\underline{a}}{}^{a}\dd \hat E_{\underline{b} a})\otimes \mathds{1}_{2}\nonumber\\
    &-\frac18e^{-\frac{\phi}{4}}\hat\gamma^{\frac12}\left(\gamma \check{\not{\mathcal{F}}}{}^{\underline a}_{(2)}\otimes\Gamma_{\underline a} -2\gamma^{\underline\rho}\otimes\Gamma_{\underline a}\check{\not{\mathcal{F}}}{}^{\underline a}_{\underline{\mu\rho}}e^{\underline\mu}\right)\otimes\ri\sigma_2\nonumber\\
    &+\frac{1}{16}e^{-\frac{\phi}{4}}\left(\gamma\not H_{(2)\underline{a}}\otimes\Gamma_*\Gamma^{\underline{a}}-4\gamma^{\underline{\rho}}\otimes\Gamma_*\Gamma^{\underline{a}}H_{\underline{\mu\rho a}}e^{\underline{\mu}}\right)\otimes \sigma_{3}\nonumber \\
    &-\frac{1}{16}e^{\frac{\phi}{2}}(\gamma\not F_{(3)}-2\gamma^{\underline{\rho\lambda}}F_{\underline{\mu\rho\lambda}}e^{\underline{\mu}})\otimes\mathds{1}_{4}\otimes\sigma_{1}-\frac{1}{32}\hat\gamma^{\frac12}\left(\gamma\not F_{\underline{ab}}\otimes \Gamma^{\underline{ab}}-4F_{\underline{\mu ab}}\mathds{1}_8\otimes\Gamma^{\underline{ab}}e^{\underline\mu}\right)\otimes\sigma_1\,,
\end{align}
with $\Delta$ defined as in \eqref{defDelta} and $\gamma=\gamma_\mu \dd x^\mu$, $\gamma_\mu$ being the $D=6$ gamma-matrices, defined in Appendix \ref{appA}. The expressions \eqref{susylambda}, \eqref{susypsia}, \eqref{susypsimu} need to vanish simultaneously for any bosonic solution. In particular, the vanishing of \eqref{susypsimu} implies the following integrability equation
\begin{align}
    (\dd\mathbf{W}+\mathbf{W}\wedge \mathbf{W})\,\boldsymbol{\epsilon}=0\,.
\end{align}
Finally, the consistency of the chosen truncation of type IIB supergravity requires the supersymmetry variation of $C_0$ to be zero, while the one of the dilaton needs to remain non-vanishing. This is consistently achieved by requiring the following conditions on the fermions 
\begin{align}
    \Gamma_{*4}\sigma_{1}\boldsymbol{\epsilon} & =\boldsymbol{\epsilon}\,,\qquad \Gamma_{*4}\sigma_{1}\boldsymbol{\lambda}  =-\boldsymbol{\lambda}\,, \nonumber\\
     \Gamma_{*4}\sigma_{1}\boldsymbol{\Psi}_\mu & =\boldsymbol{\Psi}_\mu\,, \qquad \Gamma_{*4}\sigma_{1}\boldsymbol{\Psi}_a =-\boldsymbol{\Psi}_a\,.
\end{align}

\section{Review of warped three-dimensional spaces }
In the next section, we will present various novel solutions of the six-dimensional theory mentioned earlier and discuss their uplift to type IIB supergravity. Before providing the explicit expressions of the backgrounds, let us introduce, in the present section, the building blocks that will be used to construct them, namely the three-dimensional warped anti-de Sitter ${\rm WAdS}_3$ and the warped 3-sphere ${\rm WS}^3$, giving a unified, though concise, description of them. \par
As pointed out in the Introduction, three-dimensional warped anti-de Sitter spacetimes, and orbifolds thereof, were extensively studied within TMG \cite{Anninos:2008fx,Chow:2009km,Deger:2013yla} (see also \cite{Bieliavsky:2024hus} for a recent, in-depth review of warped-AdS solutions). They also occur as solutions to new massive gravity (NMG) \cite{Clement:2009gq,Corral:2024xfv}, minimal massive gravity (MMG) \cite{Setare:2017xlu,Nam:2018gju,Deger:2024dbz}, as well as three-dimensional (super-) gravity with torsion \cite{Andrianopoli:2023dfm,Andrianopoli:2024twc}.\par
A warped anti-de Sitter spacetime as well as a warped 3-sphere, are instances of K-contact, $\eta$-Einstein manifolds \cite{Okumura,Boyer:2004eh}, with Lorentzian and Euclidean metrics $g_3$, respectively. $\eta$-Einstein manifolds are deformations of Einstein manifolds, generated by a suitable 1-form field $\mathsf{A}$, endowed with a normal contact structure. These geometries are characterized by the following general expression of their Ricci tensor:
\begin{align}
    R_{3}=a\, g_{3}+ b\, \mathsf{A}\otimes\mathsf{A}\,, \label{Ricci = g + AA}
\end{align}
with constants  $a,b$. The 1-form $\mathsf{A}$ characterizing this geometry satisfies a chiral, linear, differential, Beltrami-like equation (see \cite{Andrianopoli:2023dfm} for a more detailed discussion):
\begin{align}
    \star_{g_{3}}\, d \mathsf{A}= m \mathsf{A}\,, \label{beltrami eq}
\end{align}
with a constant parameter $m$. Eq. \eqref{beltrami eq} is the ``square root'' of the Proca equation for a massive 1-form field in three dimensions \cite{Townsend:1983xs}. The sign of $m$ defines the \emph{chirality} of this field and determines which of the two factors in the isometry group of the undeformed space, which is ${\rm SL}(2,\mathbb{R})_L\times {\rm SL}(2,\mathbb{R})_R$ for ${\rm AdS}_3$ and ${\rm SU}(2)_L\times {\rm SU}(2)_R$ for ${\rm S}^3$, is broken. The relative chirality of the warpings of the ${\rm WAdS}_3$ and ${\rm WS}^3$ factors affects general properties of the six-dimensional background, such as supersymmetry  \cite{El-Showk:2011euy}. If the vector field $\xi$, dual to $\mathsf{A}$, is a Killing vector, the geometry is named K-contact, and the norm of $\xi$ can be shown to be constant, characteristic of the space. These deformations exist both in Euclidean and Lorentzian signature: while in the former case, the norm of the vector $\xi$ is always positive, there exist three kinds of deformations in the latter, depending on whether $\xi$ is lightlike, timelike or spacelike. In what follows, we will describe deformations of the 3-sphere and $\rm AdS_3$ presenting the above-mentioned features. These three-dimensional spaces are instrumental in constructing solutions to the six-dimensional theory and, consequently, new solutions to IIB supergravity. 

\subsection{Euclidean deformation} In the Euclidean case, we will consider the warped 3-sphere $\rm WS^{3}$, with the deforming parameter $\Omega$. We describe it in terms of the coordinates $(\psi,\varphi_{1},\varphi_{2})$, with metric  
\begin{align}\label{euclideanmetric}
\dd s^{2}(\rm WS^{3}) & =\ell^{T}\ell \,, \qquad \ell =(\ell^1,\ell^2,\ell^3) \, ,
\end{align}
and vielbein given by
\begin{align}
\ell^{1} &=\frac{1}{2}e^{\frac{\Omega}{4}}\dd\psi\,,\hspace{0.8cm}\ell^{2}=\frac{1}{2}e^{\frac{\Omega}{4}}\sin\psi\dd\varphi_{1}\,,\hspace{0.8cm}\ell^{3}=\frac{1}{2}e^{-\frac{3\Omega}{4}}(\dd\varphi_{2}-\cos\psi\dd\varphi_{1})\,,
\end{align}
where the coordinate ranges are $\psi\in [0,\pi],\,\varphi_1\in [0,2\pi),\,\varphi_2\in [0,4\pi)$.
The volume form is defined as $\Vol(\rm W{S}^{3})= \ell^0 \wedge \ell^1 \wedge \ell^2$, while the Ricci tensor reads as in \eqref{Ricci = g + AA}, with 
\begin{align}\label{vecS3}
    \mathsf{A}_\odot= \dd\varphi_2 - \cos \psi\dd \varphi_1\,,  
\end{align}
and $a=2e^{-\frac{5 \Omega }{2}} \left(2 e^{2 \Omega }-1\right)$, $b=-e^{-4 \Omega } (e^{2 \Omega }-1)$. The Beltrami equation is satisfied with $m=2 e^{-\frac{5 \Omega }{4}}$. \\
The Killing vectors of this geometry are
\begin{align*}
\xi_{1} & =-\cos\varphi_{1}\partial_{\psi}+\sin\varphi_{1}(\csc\psi\partial_{\varphi_{2}} + \cot \psi \partial_{\varphi_{1}})\,,\qquad\xi_{2}= \sin\varphi_{1}\partial_{\psi} + \cos\varphi_{1}(\csc\psi\partial_{\varphi_{2}}+\cot\psi\partial_{\varphi_{1}})\,,\\
\xi_{3} & =\partial_{\varphi_{1}}\,,\qquad\xi_{4}=\partial_{\varphi_{2}}\,.
\end{align*}
The vectors $\{\xi_1,\xi_2,\xi_3\}$ close into an $\mathfrak{ so}(3)$ algebra $[\xi_i,\xi_j]=\epsilon_{ijk}\xi_k$, and $\xi_4$ is a $\mathfrak u(1)$ factor. These are the only non-vanishing commutators among the four Killing vectors. Then, the isometry group of the space is $\rm U(2)$, which is reminiscent of the fact that the space is a fiber bundle of the form $\rm S^1 \hookrightarrow \rm WS^3 \stackrel{\pi}{\to} S^2$. In the limit $\Omega\to0$ the space becomes the round 3-sphere, described in the coordinates of the Hopf fibration. Note that the vector dual to $\mathsf{A} _\odot$ is proportional to the Killing vector $\xi_4$, leading to the following identities in the Lie derivative of the 1-form: $\mathcal{L}_{\xi_{i}} \mathsf{A}_\odot = \mathcal{L}_{\xi_{i}} \dd \mathsf{A}_\odot = 0$, $\forall i=1,\dots,4.$ This implies that any object constructed from $\mathsf{A}_\odot$ will be invariant under $\rm U(2)$. Warped 3-spheres as gravitational instantons have been shown to be relevant in the context of AdS/CFT in both supersymmetric and non-supersymmetric cases (see \cite{Martelli:2013aqa,Bobev:2016sap,Bueno:2018yzo, Canfora:2023bug} and references therein).

\subsection{Lightlike deformation} \label{section lightlike 3D}

The lightlike Lorentzian cousin of the geometry discussed above corresponds to warped $\rm AdS_3$ that we denoted by $\mathrm{WAdS}_{3}^{0}$. The deformation is controlled by a parameter $\alpha$ and the $\rm AdS_3$ limit is recovered when $\alpha=0$. We will consider coordinates $(x_{-},u,x_{+})$, in terms of which the metric reads
\begin{align}
    \dd s^{2}(\mathrm{WAdS}_{3}^{0}) & =\mathtt{e}^{T}_{\times}\eta_{3}\mathtt{e}_{\times} = \frac{\dd u^2+\dd x_{+}\dd x_{-} }{u^2}-\alpha \frac{\dd x_{-}^2}{u^4} \,, \qquad \mathtt{e}_{\times} = (\mathtt{e}^0_{\times},\mathtt{e}^1_{\times},\mathtt{e}^2_{\times}) \, , \label{metric WAdS3 null}
\end{align}
with vielbein
\begin{align}  \label{metric AdS0}
   \mathtt{e}^{0}_{\times} & =\dd x_{+}-(1+4\alpha)\frac{\dd x_{-}}{4u^{2}}\,, \qquad   \qquad  \mathtt{e}^{1}_{\times}=\frac{\dd u}{u}\,, \qquad \hspace{.5cm}\mathtt{e}^{2}_{\times}=\dd x_{+}+(1-4\alpha)\frac{\dd x_{-}}{4u^{2}}\,,
\end{align}
and flat metric $\eta_{3}=\mathrm{diag}(-1,1,1)$. The volume form is defined as $\Vol(\rm WAdS_3^0)= e^0_\times \wedge e^1_\times \wedge e^2_\times$. \\
In the $\alpha \to 0$ limit, the space becomes $\rm AdS_3$ in the Poincar\'e patch. Indeed $u\in [0,+\infty)$ is the radial coordinate with asymptotic region at $u=0$ and origin at $u \to \infty$, whereas the coordinates $x_+,x_- \in \bR$ are two null coordinates. 

Turning on $\alpha$, the space becomes $\rm WAdS_3^0$, as the Ricci tensor can be written as \eqref{Ricci = g + AA} in terms of the null 1-form  
\begin{align}\label{vecWads0} 
    \mathsf{A}_\times =  \frac{\dd x_-}{u^2}\, ,
\end{align}
and constants $a=-2$, $b=4 \alpha$, and the mass of the Beltrami equation \eqref{beltrami eq} is $m=-2$. In the region close to the origin $u\to \infty$, namely in the IR, the space tends to $\rm AdS_3$. The asymptotic region is still located at $u \to 0$, but the presence of $\alpha$ changes the asymptotic structure, as the leading term at infinity in the component $g_{x_{-} x_{-}}$ is $-\alpha u^{-4}$. The vector $\partial/\partial x_-$ is a timelike vector field and tends to be null in the limit $u\to \infty$, while $\partial/\partial x_+$ remains null. The deformed space has an anisotropic rescaling symmetry of the coordinates given by $(x_-,u,x_+)\to (\lambda^2 x_-,\lambda u,\lambda^{-1} x_+)$ which is present in the asymptotic region, and corresponds to the dynamical exponent $n=2$ in \cite{Bobev:2011qx}. This anisotropic rescaling symmetry suggests that, in the UV, the dual field theory is non-relativistic.

The Killing vectors of $\rm WAdS_3^0$ are
\begin{align}
\xi_{1} & =x_{-}^{2}\partial_{x_{-}}+x_{-}u\partial_{u}-u^{2}\partial_{x_{+}}\,,\qquad\xi_{2}=2x_{-}\partial_{x_{-}}+u\partial_{u}\,,\qquad\xi_{3}=\partial_{x_{-}}\,,\qquad\xi_{4}=\partial_{x_{+}}\,,\nonumber \\
 & [\xi_{2},\xi_{1}]=2\xi_{1}\,, \qquad [\xi_{2},\xi_{3}]=-2\xi_{3}\,, \qquad [\xi_{1},\xi_{3}]=-\xi_{2} \, .
\end{align}
The vectors $\{ \xi_1,\xi_2,\xi_3 \}$ close a $\frak{sl}_2(\bR)$ algebra, whereas $\xi_4$ generates a $\frak{u}(1)$ factor: the isometry group therefore is $\rm SL(2,\bR)\times U(1)$. It is interesting to note that the dual vector to the 1-form $\mathsf{A}_\times $ is the Killing vector $\xi_4$, that in addition to be Killing is also tangent to a geodesic, implying that the metric \eqref{metric AdS0} is written in Kerr-Schild form: $\dd s^2(\rm AdS_3) - \alpha \mathsf{A}_\times \otimes \mathsf{A}_\times$.

\subsection{Spacelike deformation} 

The spacelike deformation of $\rm AdS_3$ will be referred to as $\mathrm{WAdS}_{3}^{+}$ and the parameter controlling the deformation is dubbed $\omega$. This space will be described in terms of $(t,\rho,\theta)$ coordinates, with $\rho\geq 0$, $t\in \mathbb R$ and $\theta \sim \theta+2\pi$, in terms of which the metric reads
\begin{align}\label{spacelikemetric}
    \dd s^{2}(\mathrm{WAdS}_{3}^{+}) & =\mathtt{e}_\to^{T}\eta_{3}\mathtt{e}_\to\,, \qquad \mathtt{e}_\to = (\mathtt{e}^0_\to, \mathtt{e}^1_\to,\mathtt{e}^2_\to) \, ,
\end{align}
with vielbein 
\begin{align}
    \mathtt{e}^0_{ \to}=\frac{e^{\frac{\omega}{4}}}{2} \sinh{\rho}  \,\dd t \,, \qquad \mathtt{e}^1_\to=\frac{e^{\frac{\omega}{4}}}{2}\,\dd \rho \,,\qquad \mathtt{e}^2_\to=\frac{e^{-\frac{3\omega}{4}}}{2} (\dd\theta + \cosh{\rho} \dd t) \, .
\end{align}
The volume form is denoted by $\Vol (\rm WAdS_3^+)=\mathtt e^0_\to\wedge \mathtt e^1_\to\wedge \mathtt e^2_\to$. 

This spacetime has the structure of a fiber bundle $S^1\hookrightarrow \rm WAdS_3^+\stackrel{\pi}{\to} \rm AdS_2$, where the base manifold is $\rm AdS_2$ with coordinates $(t,\rho )$. Moreover, the deformation is generated by the following spacelike vector field
\begin{align}\label{vecWadsp}
    \mathsf{A}_\to = \dd \theta +\cosh \rho \dd t\,,
\end{align}
which deforms the Einstein structure with constants $a=-2e^{-\frac{5 \omega }{2}} \left(2 e^{2 \omega }-1\right)$, $b=e^{-4\omega}(e^{2 \omega}-1)$. In addition, it satisfies the Beltrami equation in \eqref{beltrami eq} with $m=2 e^{-\frac{5 \omega }{4}}$. 

In this case, the Killing vectors are the ones of $\rm AdS_2$, in addition to the Killing vector associated with translation in $\theta$, leading to the isometry group $\rm SL(2, \bR)\times \rm U (1)$. Explicitly, the Killing vectors and the corresponding non-vanishing Lie brackets are 
\begin{align*}
\xi_{1} & =e^{t}(\partial_{\rho}-\coth\rho\partial_{t}+\mathrm{csch}\rho\partial_{\theta})\,,\qquad\xi_{2}=e^{-t}(\partial_{\rho}+\coth\rho\partial_{t}-\mathrm{csch}\rho\partial_{\theta})\,,\qquad\xi_{3}=\partial_{t}\,,\qquad\xi_{4}=\partial_{\theta}\,, \\
&\quad [\xi_{1},\xi_{2}] = 2\xi_{3}\,, \qquad [\xi_{3},\xi_{2}]=-\xi_{2}\,, \qquad[\xi_{3},\xi_{1}]=\xi_{1} \, .
\end{align*}

\subsection{Timelike deformation} 

The timelike deformation of $\rm AdS_3$ will be denoted by $\mathrm{WAdS}_{3}^{-}$. The metric, in terms of $(t,\rho,v)$ coordinates, with $\rho\geq0$ and $t,v\in \bR$, reads
\begin{align}\label{timelikemetric}
    \dd s^{2}(\mathrm{WAdS}_{3}^{-}) & =\mathtt{e}_\uparrow^{T}\eta_{3}\mathtt{e}_\uparrow\,, \qquad \mathtt{e}_\uparrow = (\mathtt{e}^0_\uparrow, \mathtt{e}^1_\uparrow,\mathtt{e}^2_\uparrow) \, ,
\end{align}
with vielbein
\begin{align}
{\tt e}_{\uparrow}^{0}= \frac{e^{-\frac{3\omega}{4}}}{2}(\dd t+\cosh\rho\dd v)\,,\hspace{1.2cm}
{\tt e}_{\uparrow}^{1}= \frac{e^{\frac{\omega}{4}}}{2}\dd\rho\,,\hspace{1.2cm} 
{\tt e}_{\uparrow}^{2}= \frac{e^{\frac{\omega}{4}}}{2}\sinh\rho\dd v\,.
\end{align}
In the limit $\omega \to 0$, the space becomes $\rm AdS_3$ in global coordinates, as it can be seen from the embedding in $\bR^{2,2}$ satisfying $-X_0^2+X_1^2+X_2^2-X_3^2=-1$ with
\begin{align}
    X_0 =  \cosh\frac{\rho}{2} \cos \frac{v+t}{2}  \, , \hspace{0.1cm} X_1 =  \sinh\frac{\rho}{2} \cos \frac{v-t}{2} \, , \hspace{0.1cm}
    X_2 = \sinh \frac{\rho}{2} \sin \frac{v-t}{2} \, , \hspace{0.1cm} X_3 = \cosh \frac{\rho}{2} \sin \frac{v+t}{2} \, . 
\end{align}
The pullback of the $\bR^{2,2}$ metric leads to \eqref{timelikemetric} with $\omega=0$. Let us observe that, in these coordinates, $\partial_t$ and $\partial_v$ are both timelike vectors with constant norm, while $\partial_\rho$ is spacelike.

This spacetime is a fiber bundle $\bR \hookrightarrow \rm{WAdS_3^-} \stackrel{\pi}{\to} \mathds{H}^2$ with base manifold being the two-dimensional hyperbolic space $\mathds{H}^2$ and with $\mathbb R$ as fiber, parameterized by the time coordinate.
The Ricci tensor can be expressed as in \eqref{Ricci = g + AA}, with constants $a=2e^{-\frac{5\omega}{2}}(1-2e^{2\omega})$, $b=-e^{-4\omega}(e^{2\omega}-1)$ and vector
\begin{align}\label{vecWadsm}
    \mathsf{A}_{\uparrow}=\dd t + \cosh\rho\dd v \,.
\end{align} 
The Beltrami equation \eqref{beltrami eq} is satisfied with mass parameter $m=2 e^{-\frac{5 \omega}{4}}$.

At last, the Killing vectors of this space and their non-vanishing Lie bracket are
\begin{align}
\xi_{1} & =\cos{v}(\coth\rho\partial_{v}-\mathrm{csch}\rho\partial_{t})+\sin{v} \partial_{\rho}\,,\quad
\xi_{2}=\sin{v}(-\coth{\rho}\partial_{v}+\mathrm{csch}\rho\partial_{t}) +\cos{v}\partial_{\rho}\,,\\ \qquad
\xi_{3} &= \partial_{\theta}\,,\quad
\xi_{4}=\partial_{t}\,,\nonumber \\ & [\xi_{1},\xi_{2}]=-\xi_{3}\,,\qquad[\xi_{3},\xi_{1}]=\xi_{2}\,,\qquad[\xi_{2},\xi_{3}]=\xi_{1}\,.
\end{align}
The vector fields $\{\xi_1,\xi_2,\xi_3\}$ close a $\frak{so}(1,2)$ algebra, which is the algebra of the isometry group of the base manifold. Therefore the isometry group of the space is $\rm SO (1,2)\times U(1)$. Note that the dual vector to $\mathsf{A}_{\uparrow}$ is proportional to $\xi_4$, which in turn commutes with all Killing vectors, leading to the following identity involving the Lie derivative $\mathcal{L}_{\xi_i} \mathsf{A}_{\uparrow}=\mathcal{L}_{\xi_i} \dd \mathsf{A}_{\uparrow} = 0$, $\forall i=1,\dots, 4$.

\section{Double-deformed lightlike solutions}
In this section, we present a new 3-parameter solution of the six-dimensional theory and describe its uplift to ten dimensions. The spacetime is a direct product between a null warped $\rm AdS_3$, with the deformation controlled by the parameter $\alpha\geq0$, and a warped 3-sphere, with deformation parameter $\Omega\geq0$. The six-dimensional metric is of the form
\begin{align}\label{sol6dlight}
\dd s^2 = \dd s^2({\rm WAdS_3^0}) + e^{-\frac{\Omega}{2}} \dd s^2 ({\rm WS^3}) \, ,
\end{align}
where the metric and vielbein of the $\mathrm{WAdS}^0_{3}$ and $\rm WS^3$ are given in \eqref{metric WAdS3 null} and in \eqref{euclideanmetric}, respectively. We consider the orientation of the
six-dimensional manifold to be
$\epsilon_{x_{+} u x_{-} \psi \varphi_{1} \varphi_{2}}=1$. It is convenient to introduce the following 1-forms
\begin{align}
\mathsf{A}_{1}  =\frac{2\sqrt{\alpha}\lambda}{\sqrt{1+3e^{2\Omega}}}\mathsf{A}_\times \,,\hspace{1.5cm}
\mathsf{A}_{2}  = e^{-\Omega} \lambda \mathsf{A}_\odot \,.
\end{align}
As shown in the previous section, $\mathsf{A}_\times$ and $\mathsf A_\odot$ support the deformation of $\rm WAdS_3^0$ and $\rm WS^3$, respectively. Moreover, they allow the introduction of higher-degree forms, which are instrumental for constructing solutions to the equations of motion. \\
The following bosonic configuration with non-vanishing 3-form field strength, six-dimensional dilaton and vector fields is a solution of the six-dimensional equations of motion \eqref{eqmotion1}, \eqref{eqmotion2}:
\begin{align}\label{F3 null case}
F_{(3)} & =\frac{1}{\lambda^{2}}\left[- \Vol(\mathrm{WAdS}_{3}^{0}) - e^{\frac{\Omega}{4}}\Vol({\rm WS}^{3}) + \frac{e^{2\Omega}}{4\lambda^{2}}\dd (\mathsf{A}_{1} \wedge  \mathsf{A}_{2} ) \right]\,, \nonumber\\
e^{\phi/2} & =2\lambda^{2}e^{-\Omega}\,,\nonumber\\
\check{\mathcal{F}}^{a}_{(2)} & =H_{(2)a}=\sqrt{\sinh\Omega}(\dd \mathsf{A}_{1} + \dd \mathsf{A}_{2})\delta_{a}^{4}\,.
\end{align}
The solution is real and supersymmetric for $\lambda\neq0$ and $\Omega,\alpha\geq0$ and the corresponding common solution to all Killing spinor equations reads
\begin{align} \label{KS double null deformation}
\epsilon(\psi,\varphi_{1}) & = \frac{1}{8} e^{-\frac{\psi}{2}\gamma^{45}\gamma_{*}} e^{+\frac{\varphi_{1}}{2} \gamma^{34}} (\mathds{1}+\Gamma_{*}\sigma_{1}) (\mathds{1}-\gamma^{02}) \left[\mathds{1}-\sqrt{e^{2\Omega}-1}\gamma^{5}\Gamma^{4}\ri\sigma_{2} + e^{\Omega}\gamma_{*}\right]\bP_{10}\epsilon_{0}\,,
\end{align}
where we are using the Clifford algebra basis presented in Appendix \ref{appA}. The constant complex spinor $\epsilon_0$ satisfies the Majorana condition \eqref{pseudo-majorana condition}, which halves the number of independent components. The background preserves 4 supercharges and the Killing spinor depends on the coordinates $(\psi,\varphi_1)$ of the base manifold $\rm S^2$  of the Hopf fibration of $\rm WS^3$ in \eqref{euclideanmetric}. The Killing spinor is globally defined, antiperiodic in $\varphi_1$, and has a smooth limit when the deformation of the 3-sphere goes to zero $\Omega \to 0$ or when the warping of $\rm AdS_3$ vanishes $\alpha \to 0$. In each of these limits, there is an enhancement of the isometry group, but only in the latter case is there a doubling of the number of preserved supercharges.

In the limit $\Omega \to 0$, the space acquires the form of $\rm WAdS_3^0\times \rm S^3$, the six-dimensional vectors vanish, and all the backreaction of the warping of $\rm AdS_3$ is supported by the 3-form field strength. The background still preserves 4 supercharges, and the Killing spinor is the same as \eqref{KS double null deformation}. The isometry group is enhanced to $\rm SL(2,\bR) \times \rm U(1) \times \rm SO(4)$, but the presence of the term containing the 1-forms $\mathsf{A}_\times, \mathsf{A}_\odot$ in the 3-form field strength, reduces the symmetry of the background to be $\rm SL(2,\bR) \times \rm U(1) \times \rm U(2)$. In this limit, our configuration can in principle be compared with the one in \cite{Deger:2024xnd}: however, \eqref{sol6dlight} is a static solution, whereas the configuration in the reference under consideration is stationary and does not posses a limit in which the rotation is absent and the deformation of $\rm AdS_3$ is still present.

Finally, in the $\alpha \to 0$ limit, the spacetime becomes $\rm AdS_3\times \rm WS^3$ and corresponds to the BPS configuration constructed in \cite{Eloy:2021fhc}. In this case, the isometry group is enhanced to $\rm SL(2,\bR)\times SL(2,\bR) \times U(2)$, and the background preserves 8 supercharges. The most general Killing spinor in this limit is $$ \mathds{M}(x_+)e^{\frac{\psi}{2} \gamma^{45}}e^{-\frac{\varphi_1}{2} \gamma^{34}}(\mathds{1}+\Gamma_{*}\sigma_{1}) \left[\mathds{1}-\sqrt{e^{2\Omega}-1}\gamma^{5}\Gamma^{4}\ri\sigma_{2} + e^{\Omega}\gamma_{*}\right]\bP_{10}\epsilon_{0}\,,$$
where $\mathds{M}(x_+)$ is an invertible matrix encoding the dependence of the spinor on the $\rm AdS_3$ coordinate $x_+$.

Following the uplift formulas in \eqref{uplift type IIB}, the null deformation configuration discussed above can be embedded in type IIB supergravity as
\begin{align}
\dd s_{10}^{2} & =\frac{e^{\frac{\Omega}{2}}}{\sqrt{2}\lambda} [ \dd s^2({\rm WAdS_3^0}) + e^{-\frac{\Omega}{2}} \dd s^2 ({\rm WS^3})  ]+\sqrt{2}e^{-\frac{\Omega}{2}}\lambda\dd y^{a}\dd y^{a}\,,\nonumber \\
\mathsf{F}_{5} & =\sqrt{\sinh\Omega}\left[\dd y^{1}\wedge\dd y^{2}\wedge\dd y^{3} \wedge \dd  + \lambda^{-2} (\mathrm{Vol}(\mathrm{WAdS}_{3}^{0})-e^{\frac{\Omega}{4}}\Vol(\mathrm{WS}^{3})) \wedge \dd y^{4} \wedge \right] (\mathsf{A}_1 + \mathsf{A}_2)\,, \nonumber \\
\mathsf{F}_{3} & =\frac{1}{\lambda^{2}}\left[-\mathrm{Vol}(\mathrm{WAdS}_{3}^{0})-e^{\frac{\Omega}{4}}\Vol(\mathrm{WS}^{3})+\frac{e^{2\Omega}}{4\lambda^{2}}\dd(\mathsf{A}_{1}\wedge\mathsf{A}_{2})\right]\,,\nonumber \\
\mathsf{F}_{1} & =0\,,\nonumber \\
e^{\Phi} & =2\lambda^{2}e^{-\Omega}\,,\nonumber \\
H_{3} & =\sqrt{\sinh{\Omega}}\dd(\mathsf{A}_{1}+\mathsf{A}_{2})\wedge\dd y^{4}\,,
\end{align}
where we are following the conventions presented in Appendix \ref{appB}. The ten-dimensional spacetime contains compact submanifolds that allow us to compute the supergravity charges: in particular, we use ${\rm WS}^3,\,  T^4,\,  T^3$ with coordinates $y^1,\, y^2,\, y^3$, and $S^2\subseteq S^3$. The supergravity charges are
\begin{align}
    Q_{\rm D1} = -2 e^{-3 \Omega} \lambda^2 \, , \quad Q_{\rm D3} = \frac12 e^{-\Omega} \lambda \sqrt{\sinh \Omega} \, , \quad Q_{\rm D5} =-\frac{1}{2 \lambda^2} \, , \qquad Q_{\rm NS5} = e^{-\Omega} \lambda \sqrt{\sinh \Omega} \, .
\end{align}
Therefore, we argue that a putative microscopic description of this solution should contain at least the brane structure shown in Table \ref{table null background}.

\begin{table}[]
    \centering
\begin{tabular}{c|cccccccccc}
 & $x_{-}$ & $x_{+}$ & $u$ & $\psi$ & $\varphi_{1}$ & $\varphi_{2}$ & $y^{1}$ & $y^{2}$ & $y^{3}$ & $y^{4}$\tabularnewline
\hline 
D1 & $\times$ & $\times$ &  &  &  &  &  &  &  & \tabularnewline
D3 & $\times$ & $\times$ &  &  &  & $\times$ &  &  &  & $\times$\tabularnewline
D5 & $\times$ & $\times$ &  &  &  &  & $\times$ & $\times$ & $\times$ & $\times$\tabularnewline
NS5 & $\times$ & $\times$ &  &  &  & $\times$ & $\times$ & $\times$ & $\times$ & \tabularnewline
\end{tabular}
    \caption{A possible brane setup for the $\mathrm{WAdS}^0_{3}\times \rm WS^3$ solution.}
    \label{table null background}
\end{table}

\section{Double-deformed spacelike solutions}
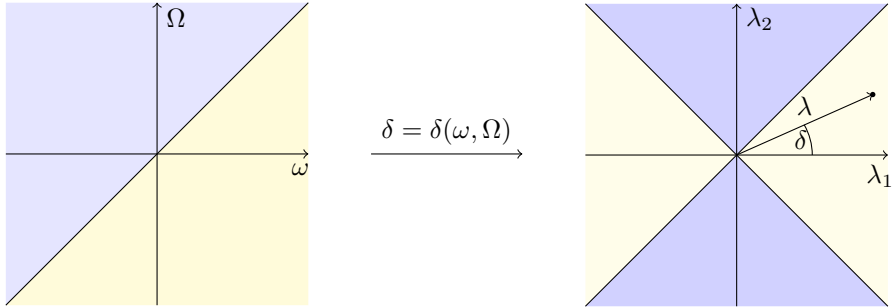
\begin{figure}[h]

\begin{minipage}{.07\textwidth}
    \,
\end{minipage}
\begin{minipage}{.5\textwidth}
    \centering
    
    \begin{tikzpicture}
    
    \fill[yellow!20,opacity=0.9] (-2,-2) -- (2,-2) -- (2,2) -- cycle;
    \fill[blue!20,opacity=0.5] (-2,-2) -- (-2,2) -- (2,2) -- cycle;
    
    \draw (-2,-2)--(2,2);
    \draw[->] (-2,0)--(2,0) node[below,xshift=-.1cm] {$\omega$};
    \draw[->] (0,-2)--(0,2) node[right,yshift=-0.2cm] {$\Omega$};

    \draw[->] (6-1.5-1.67,0)--(6+0.5-1.67,0) node[midway,above] {$\delta = \delta(\omega,\Omega)$};
    \end{tikzpicture}
\end{minipage}
\begin{minipage}{.3\textwidth}
\centering
\begin{tikzpicture}
     \fill[yellow!20,opacity=0.5] (10,0) -- (12,-2) -- (12,2) -- cycle;
     \fill[yellow!20,opacity=0.5] (10,0) -- (8,-2) -- (8,2) -- cycle;
    \fill[blue!20,opacity=0.9] (10,0) -- (-2+10,2) -- (12,2) -- cycle;
    \fill[blue!20,opacity=0.9] (10,0) -- (-2+10,-2) -- (12,-2) -- cycle;
    \draw[->] (-2+10,0)--(2+10,0) node[below,xshift=-.1cm] {$\lambda_1$};
    \draw[->] (10,-2)--(10,2) node[right,yshift=-0.2cm] {$\lambda_2$};
    
    \draw (10-2,-2)--(10+2,2);
    \draw (10+2,-2)--(10-2,2);
    \fill (11.8,0.8) circle (1.pt);
    \draw[->] (10,0)--(11.8,0.8) node[midway, above] {$\lambda$};
    \draw (10+1.0,0) arc[start angle=0, end angle=30, radius=0.8cm];
    \node at (10+1-0.15,0.2) {$\delta$};
\end{tikzpicture}
\end{minipage}
\begin{minipage}{.1\textwidth}
    \,
\end{minipage}

\caption{Relation between the space of deformation parameters $(\omega,\Omega)$ and solutions $\delta$ of the constraint \eqref{spacelike constraint}. A point in the right panel is of the form $(\lambda_1,\lambda_2)=(\lambda \cos \delta,\lambda \sin \delta)$, where $\lambda\geq0$. Given a point in the blue (yellow) region in the plane $(\omega,\Omega)$, solutions of the constraint lie in the blue (yellow) region in the plane $(\lambda_1,\lambda_2)$.}
    \label{fig: map delta}

\end{figure}

The purpose of this section is providing a new solution of both six and ten-dimensional theories, involving spacelike warped $\mathrm{AdS_3}$, $\mathrm{WAdS_3^+}$, and the warped sphere $\rm W S^3$. The six-dimensional metric is of the form $\mathrm{WAdS}_{3}^{+}(\omega)\times \rm W{S}^{3}(\Omega)$
and it is given by
\begin{align}
    \dd s^2 = e^{\frac{\Omega}{2}}\dd s^2({\rm WAdS_3^+}) + e^{\frac{\omega}{2}}\dd s^2 ({\rm WS^3}) \, ,
\end{align}
with three-dimensional metrics given in \eqref{spacelikemetric}, \eqref{euclideanmetric} and orientation $\epsilon_{t \rho \theta \psi \varphi_1, \varphi_2} =1$. As in the previous case, it is convenient to define the following 1-forms
\begin{align}\label{vecspace}
\mathsf{A}_{1}  =\lambda\cos\delta\,\mathsf{A}_{\rightarrow}\,, \hspace{1cm}
\mathsf{A}_{2}  =\lambda\sin\delta\,\mathsf{A}_{\odot}\,,
\end{align}
where $\lambda,\delta$ are real constants. Let us then consider the following bosonic configuration involving the 3-form field strength, the six-dimensional dilaton and the vectors
\begin{align}
F_{(3)} & =\frac{e^{-\omega-\Omega}}{\lambda^{2}\cos2\delta}\left[e^{\frac{\omega}{4}}\Vol(\mathrm{WAdS}_{3}^{+})+e^{\frac{\Omega}{4}}\Vol(\rm WS^{3}) -\frac{1}{4\lambda^{2}}\dd(\mathsf{A}_{1}\wedge\mathsf{A}_{2})\right]\,,\nonumber\\
e^{\phi/2} & =2e^{\frac{\Omega+\omega}{2}}\lambda^{2}\,,\nonumber\\
\check{\mathcal{F}}^{a}_{(2)} & =H_{(2)a}=\sqrt{\frac{\sinh(\omega-\Omega)}{\cos 2\delta}}\,\dd(\mathsf{A}_{1}+\mathsf{A}_{2})\delta_{a}^{4}\,.
\end{align}
The Einstein equations are satisfied provided $\omega,\Omega$ and the parameter $\delta$ satisfy the following constraint

\begin{align}
    2e^{2(\omega+\Omega)}\cos^{2}2\delta+(e^{2\Omega}-e^{2\omega})\cos2\delta-(e^{2\Omega}+e^{2\omega})=0\,. \label{spacelike constraint}
\end{align}
The above equation is invariant under the change of sign of $\delta$ and under the transformation $(\omega,\Omega,\delta) \to (\Omega,\omega,\pi/2-\delta)$. The latter symmetry transformation interchanges the deformation parameters, making the configuration democratic in the two deformations parameters, which is possible as the two deformation 1-forms \eqref{vecspace} are both spacelike. Moreover, \eqref{spacelike constraint} has two real roots in the variable $\cos2\delta$ of opposite sign. 
Considering the case $\Omega>\omega$, which is depicted in blue region in the left panel of Figure \ref{fig: map delta}, one can always find a solution of the equation \eqref{spacelike constraint} for $\delta$ which leads to a real configuration. The solution of this equation will lie in the blue region in the plane $\lambda_1=\lambda \cos \delta, \lambda_2=\lambda \sin \delta$ depicted in the right panel of Figure \ref{fig: map delta}. Starting from the solution in the region $\Omega > \omega$ one can generate all the solution in the region $\Omega<\omega$ by acting with the symmetry transformations.

The configuration is supersymmetric when either $\omega$ or $\Omega$
vanish. The limit $\omega=0$ leads to the supersymmetric configuration $\rm AdS_{3}\times W{S}^{3}$, first found in \cite{Eloy:2021fhc}. The latter identifies the supersymmetric locus of the more general solution $\mathrm{AdS}_3\times \mathrm {M}^3_{\Omega, Z}$, with $\mathrm {M}^3_{\Omega, Z}$ being the $(\Omega, Z)$-family of deformed spheres introduced in the same reference. Similarly, for $\Omega=0$ the space becomes a direct product between warped $\mathrm{WAdS}_{3}^{+}$
and a round $\rm S^{3}$. In the latter limit, the solution preserves
8 supercharges and the Killing spinor is explicitly given by 
\begin{align}
\epsilon(t,\rho,\varphi_{2}) & =\frac14 e^{\frac{\rho}{2}\gamma^{02}\Gamma_{*}}e^{\frac{t}{2}\gamma^{01}}e^{\frac{\varphi_{2}}{2}\gamma^{34}}(\mathds{1}+\Gamma_{*}\sigma_{1})\left[\mathds{1}-(e^{2\omega}-1)^{\frac{1}{2}}\gamma^{2}\Gamma^{4}\ri\sigma_{2}+e^{\omega}\gamma_{*}\right]\mathds{P}_{10}\epsilon_{0}\,.
\end{align}
This is globally defined, periodic in the non-contractible coordinate $\varphi_2$ and has a well-behaved limit $\omega\to0$. The constant complex spinor $\epsilon_0$ satisfies the Majorana condition \eqref{pseudo-majorana condition} which halves the independent complex components. Similarly to the ``mirror'' solution in \cite{Eloy:2021fhc}, this configuration can be shown to correspond to the supersymmetric locus 
$\zeta^{2}+e^{-2\omega}-1=0$ of the more general solution
\begin{align}
    \dd s^{2} & = \dd s^2( {\rm LM}^3_{\omega, \zeta} )+\Xi^{-2}\dd s^2(\mathrm{S}^3)\,,\nonumber\\
    F_{(3)} & =-2 \Xi^{-1} \Vol( \mathrm{LM}^{3}_{\omega, \zeta})+2 \Vol({\rm S}^3)
    \,,\nonumber\\
    \phi & =\log\Xi^{4}\,,\nonumber\\
    \check{\mathcal {F}}^{a}_{(2)}&=H_{(2)a}  =e^{\omega}\zeta\Xi^{4}(\cosh^{2}\rho\dd\beta+\sinh^{2}\rho\dd t)\delta_{a}^{4}\,,    
\end{align}
where the three-dimensional Lorentzian metric is $\dd s^2( {\rm LM}^3_{\omega, \zeta} ) = {\tt e}^T \eta _3 {\tt e}$ with vielbeins
\begin{align}
   \mathtt e^{0} & =e^{-\omega/2}\Xi^{-1}\Sigma^{2}\sinh\rho\dd t\,, \quad \mathtt e^{1} =\Xi^{-1}\dd\rho\,,\quad \mathtt e^{2}  =e^{\omega/2}\cosh\rho\Xi^{3}(\Sigma^{-2}\dd\beta-\Sigma^{2}e^{\omega}\zeta^{2}\sinh^{2}\rho\dd t)\,,
\end{align}
and
\begin{align}
\Xi^{4} & =\frac{e^{-\omega}}{1+(\zeta^{2}+e^{-2\omega}-1)\cosh^{2}\rho}\,, \hspace{1 cm}
\Sigma^4 =\frac{e^{-\omega}}{1+(e^{-2\omega}-1)\cosh^{2}\rho}\,.
\end{align}
In this case, $\rm LM_{\omega,\zeta}^{3}$ denotes the $(\omega,\zeta)$-family of deformed $\mathrm{AdS}_3$ spaces, a Lorentzian version of the deformed spheres introduced in \cite{Eloy:2021fhc}. Finally, when both deformations vanish, the solution becomes the maximally symmetric $\mathrm{AdS}_3\times \mathrm{S}^3$.

The uplift to ten dimensions of the double-deformed spacelike solution described above reads
\begin{align}
\dd s_{10}^{2} & =e^{-\frac{\Phi}{2}}[e^{\frac{\Omega}{2}}\dd s^2({\rm WAdS_3^+}) + e^{\frac{\omega}{2}}\dd s^2 ({\rm WS^3})]+e^{\frac{\Phi}{2}}\dd y^{a}\dd y^{a}\,, \nonumber\\
\mathsf{F}_{5} &=\sqrt{\frac{\sinh{(\omega\!-\!\Omega)}}{\cos2\delta}}\!\left[\dd y^{1}\!\wedge\!\dd y^{2}\!\wedge\!\dd y^{3}\! \wedge\! \dd  +\! \frac{e^{-\omega -\Omega } }{\lambda ^2} (e^{\frac{\omega}{4}}\mathrm{Vol}(\mathrm{WAdS}_{3}^{+})-e^{\frac{\Omega}{4}}\Vol(\mathrm{WS}^{3})) \!\wedge\! \dd y^{4}\! \wedge \right]\! (\mathsf{A}_1\! + \!\mathsf{A}_2), \nonumber \\
\mathsf{F}_{3} & = \frac{e^{-\omega -\Omega }}{\lambda^{2}\cos2\delta}\left[e^{\frac{\omega}{4}}\Vol(\mathrm{WAdS}_{3}^{+})+e^{\frac{\Omega}{4}}\Vol(\rm WS^{3}) -\frac{1}{4\lambda^{2}}\dd(\mathsf{A}_{1}\wedge\mathsf{A}_{2})\right]\,,\nonumber \\
\mathsf{F}_{1} & =0\,,\nonumber \\
e^{\Phi} & = 2e^{\frac{\Omega+\omega}{2}}\lambda^{2}\,, \nonumber \\
H_{3} & =\sqrt{\frac{\sinh{(\omega\!-\!\Omega)}}{\cos2\delta}}\,\dd(\mathsf{A}_{1}+\mathsf{A}_{2})\wedge\dd y^{4}\,. 
\end{align}

\section{Double-deformed timelike solutions}
In this section, we provide a new, double-deformed solution, involving the timelike $\mathrm{WAdS}_3^-$ and the warped sphere $\mathrm{WS}^3$. The metric is given by a direct product $\mathrm{WAdS}^{-}_3(\omega)\times \rm WS^3(\Omega)$ and reads
\begin{align}
    \dd s^2=e^{\frac{\Omega}{2}}\dd s^2(\mathrm{WAdS}^{-}_3)+e^{\frac{\omega}{2}}\dd s^2(\rm WS^3)\,,
\end{align}
where the three-dimensional metrics have been defined in \eqref{timelikemetric} and \eqref{euclideanmetric}. We consider the orientation of the six-dimensional manifold to be $\epsilon_{t,\rho,v,\psi,\varphi_1,\varphi_1} = 1$. Let us now introduce the following $1$-forms
\begin{align}\label{vectime}
    \mathsf{A}_1=\lambda \cos\delta\, \mathsf{A}_{\uparrow}\,,\qquad \mathsf{A}_2=\lambda \sin\delta\,  \mathsf{A}_{\odot}\,, 
\end{align}
in terms of the vector fields defined in \eqref{vecWadsm}, \eqref{vecS3} respectively. The 3-form field strength, the six-dimensional dilaton and the vector fields read
\begin{align}
F_{(3)} & =\frac{ e^{-(\omega +\Omega )}}{ \lambda ^2}\left[\cos 2\delta\left(e^{\frac{\omega}{4}}\Vol(\mathrm{WAdS}_{3}^{-})+e^{\frac{\Omega}{4}}\Vol(\rm WS^{3})\right) +\frac{1}{4\lambda^{2}}\dd(\mathsf{A}_{1}\wedge\mathsf{A}_{2})\right],\nonumber\\
e^{\phi/2} &=2 \lambda ^2 e^{\frac{\omega +\Omega }{2}}\,,\nonumber\\
\check{\mathcal{F}}^{a}_{(2)} & =H_{(2)a}= \sqrt{\sinh{(\Omega-\omega)}}\, \dd(\mathsf{A}_{1}+\mathsf{A}_{2})\delta_{a}^{4}\,.
\end{align}
As in the spacelike case, Einstein equations are satisfied if a constraint relating the two deforming parameters and $\delta$ holds. In this case, it reads 
\begin{align}\label{constrainteqtime}
    (e^{2\Omega}+e^{2\omega})\cos^22\delta-(e^{2\Omega}-e^{2\omega})\cos2\delta-2 e^{2 (\omega +\Omega )}=0\,.
\end{align}
The field configuration is real, provided that $\Omega>\omega$, which is required for the reality of the vectors, and a solution of the constraint equation \eqref{constrainteqtime} always exists. Notice that, due to the different nature of the contact structures on $\mathrm{WAdS}_3^-$ and $\rm WS^3$, timelike and spacelike respectively, the solution is not completely democratic in the two deforming parameters, differently from the previous case.

The limit $\omega=0$ again
leads to the configuration $\rm AdS_{3}\times W{S}^{3}(\Omega)$ presented in \cite{Eloy:2021fhc}, whereas for $\Omega=0$ the space becomes a direct product between $\mathrm{WAdS}_{3}^{-}$
and a round $\rm S^{3}$. In the latter limit, the solution is again supersymmetric with globally defined, $\varphi_2$-periodic Killing spinor
\begin{align}
\epsilon(\rho,v,\varphi_{2}) &= \frac14 e^{-\frac{\rho}{2}\gamma^{02} \gamma_{*}}e^{\frac{v}{2} \gamma^{12}}e^{\frac{\varphi_{2}}{2}\gamma^{34}}(\mathds{1}+\Gamma_{*}\sigma_{1})\left[\mathds{1}- (1-e^{2\omega})^{\frac{1}{2}}\gamma^{0}\Gamma^{4} 
\ri \sigma_{2} + e^{\omega}\gamma_{*}\right]\mathds{P}_{10}\epsilon_{0}\,.
\end{align}
As in the previous cases $\epsilon_0$ is a constant complex spinor satisfying the Majorana condition \eqref{pseudo-majorana condition}. Finally, when both deformations vanish, the solution becomes the known maximally symmetric $\mathrm{AdS}_3\times \mathrm{S}^3$.

The ten-dimensional uplift of the above solution can be obtained from \eqref{uplift type IIB} and reads  
\begin{align}\label{10dtimelikesol}
\dd s_{10}^{2} & =e^{-\frac{\Phi}{2}}[e^{\frac{\Omega}{2}}\dd s^2(\mathrm{WAdS}^{-}_3)+e^{\frac{\omega}{2}}\dd s^2({\rm WS}^3)]+e^{\frac{\Phi}{2}}\dd y^{a}\dd y^{a}\,, \nonumber\\
\mathsf{F}_{5} & =\sqrt{\sinh (\Omega -\omega )}\left[\dd y^{1}\!\wedge\!\dd y^{2}\!\wedge\!\dd y^{3}\! \wedge\! \dd  + \!\frac{ e^{-\omega -\Omega }}{\lambda ^2} (e^{\frac{\omega}{4}}\mathrm{Vol}(\mathrm{WAdS}_{3}^{-})-e^{\frac{\Omega}{4}}\Vol(\mathrm{WS}^{3}))\! \wedge\! \dd y^{4} \wedge \right] (\mathsf{A}_1 \!+\! \mathsf{A}_2),\nonumber\\ 
\mathsf{F}_{3} & = \frac{ e^{-\omega -\Omega } }{ \lambda ^2}\left[\cos 2\delta\left(e^{\frac{\omega}{4}}\Vol(\mathrm{WAdS}_{3}^{-})+e^{\frac{\Omega}{4}}\Vol(\rm WS^{3})\right) +\frac{1}{4\lambda^{2}}\dd(\mathsf{A}_{1}\wedge\mathsf{A}_{2})\right]\,,\nonumber \\
\mathsf{F}_{1} & =0\,,\nonumber \\
e^{\Phi} & = 2 \lambda ^2 e^{\frac{\omega +\Omega }{2}},\nonumber \\
H_{3} & =\sqrt{\sinh (\Omega -\omega )}\,\dd(\mathsf{A}_{1}+\mathsf{A}_{2})\wedge\dd y^{4}\,. 
\end{align}
Finally, let us comment on the inequivalence between the above configuration and double-deformed solution in \cite{Hoare:2022asa}: a first distinguishing feature is that, in that case, the warpings of the two $\mathrm{WAdS}_3^-$ and $\mathrm{WS}^3$ factors coincide. In this limit, which corresponds to $\Omega=\omega$ in our notation, our $H_3$ field vanishes, whereas the one in the reference under consideration does not. Moreover, the two terms in $\mathsf F_3$ have opposite six-dimensional selfdualities, whereas in  \cite{Hoare:2022asa} $\mathsf F_3$ and $H_3$ are both selfdual. We therefore conclude that there is no $\mathrm{SL}(2,\mathbb R)$ transformation relating the two configurations in the degenerate limit $\Omega=\omega$. At last, \eqref{10dtimelikesol} was obtained by uplifting a solution of the six-dimensional model, which only supports an $H_3$ field with a leg along the torus. This further ensures that \eqref{10dtimelikesol} is not a simple generalization of \cite{Hoare:2022asa}, but a fully original solution.

\section{Conclusions}
The primary outcome of the present work is the construction of new solutions, in Type IIB supergravity, of the form ${\rm WAdS}_3\times {\rm WS}^3\times T^4$, with generic warping of ${\rm WAdS}_3$. The novel feature of these backgrounds is the \emph{double warping} of the two three-dimensional factors, with two independent deformation parameters. The double-warped lightlike solutions, in particular, are supersymmetric and preserve four supercharges. On the other hand, supersymmetric solutions featuring the warping only on the non-compact anti-de Sitter factor are also known in the literature in the special case of lightlike warping \cite{Bobev:2011qx,Kraus:2011pf,Deger:2024xnd}, although BPS backgrounds with timelike warpings have been recently constructed, though only in $D=3$, $\mathcal{N}=4$ supergravity \cite{Deger:2024obg}. Here, besides the aforementioned double-warped backgrounds, we explicitly construct solutions of the form ${\rm WAdS}_3\times {\rm S}^3\times T^4$, which are supersymmetric and feature not only lightlike but also timelike and spacelike warping of the anti-de Sitter factor. From the flux content of our ten-dimensional backgrounds, we infer a possible underlying brane configuration within the Type IIB superstring theory. All solutions discussed here are also solutions to an effective $\mathcal{N}=(1,1)$, $D=6$ theory originating from a reduction of Type IIB supergravity on a $T^4/\mathbb{Z}_2$-orientifold and describing its closed string sector. In all cases, as well as in the solutions of \cite{Eloy:2021fhc}, the independent warpings of the two three-dimensional subspaces require the presence of $D=6$ vector fields, corresponding to non-trivial components of the RR 4-form $C_4$ and the Kalb-Ramond 2-form field $B_2$ along odd cycles of $T^4$. This would make these solutions, as opposed to those discussed in \cite{El-Showk:2011euy}, not extendible to the case in which the internal torus is replaced by a $K3$ space since the latter features no odd cycles. \par
As explained at the end of Section  \ref{section2}, through an $\mathrm{SO}(4,4)$-action on the solution given here, more general type IIB background with generic constant values of the $\upvarphi^s=\{\hat\gamma_{ab},C_{ab}\}$ can be constructed. The presence of these scalars, just as for the unwarped case, does not affect the structure of the $\mathrm{WAdS_3}\times \mathrm{WS^3}$ geometry. Therefore, it is natural to expect that these scalar fields should correspond, from the point of view of the holographic duality, to the analogue, in the dual $1+1$-dimensional warped-SCFT, of exactly marginal deformations in ordinary SCFT. Moreover, giving $\upvarphi^s$ a geodesic evolution along a compact direction on the boundary of ${\rm WAdS}_3$, it would be interesting to verify the possibility of constructing a \emph{warped U-Fold}, in analogy with the general construction of U-folds in \cite{Astesiano:2024gzy}.\par
Supersymmetric backgrounds  ${\rm WAdS}_3\times {\rm S}^3\times T^4$ with lightlike warping were also found as a special case of the  Schr\"odinger invariant deformations of ${\rm AdS}_3\times {\rm S}^3\times T^4$ discussed in \cite{Bobev:2011qx,Kraus:2011pf}, when the dynamical exponent is chosen to be $n=2$. It would be interesting to extend our double-warped backgrounds to Type IIB solutions of the form ${\rm Sch}_3\times_{{\rm w}} {\rm WS}^3\times T^4$.\par
Black hole geometries can be obtained as suitable quotients of
$\mathrm{WAdS}_3$ spaces, as originally discussed in \cite{Anninos:2008fx,Moussa:2008sj,Clement:2009gq}. This means that they are locally equivalent to $\mathrm{WAdS}_3$ spaces and only differ globally. The identification procedure reduces the isometry group of the manifold to $\rm U(1)\times \rm U(1)$ and requires the removal of certain points or loci from the manifold to ensure the absence of causal singularities \cite{Banados:1992gq}. The latter behave similarly to ordinary ones, and the resulting black hole geometries possess a temperature and an entropy and, for this reason, have attracted interest from the scientific community in a holographic context. It would be interesting to study the effect of this identification on the backgrounds presented here, in particular the effect on the Killing spinors, to determine whether the resulting black holes can preserve some residual supersymmetry.\par
As a final comment, let us notice that the non-supersymmetric double-warped solutions with timelike and spacelike warpings, ${\rm WAdS}^{\mp}_3\times {\rm WS}^3$, are related to supersymmetric ones by continuous parameters which describe the warping on ${\rm AdS}_3$ or  $\rm S^3$, respectively. More specifically, sending $\Omega\rightarrow 0$ we recover the supersymmetric solutions ${\rm WAdS}^{\mp}_3\times {\rm S}^3$, while, sending $\omega\rightarrow 0$ we end up with the supersymmetric ${\rm AdS}_3\times {\rm WS}^3$ backgrounds. This brings about the issue of the stability of these non-supersymmetric backgrounds, which can be addressed. This would require the computation of the full Kaluza-Klein spectrum on the backgrounds. Nevertheless, it is reasonable to expect that, since the supersymmetry breaking is induced by a continuous parameter, for sufficiently small values of the latter, the Kaluza-Klein masses can stay above the threshold value for stability. Such an analysis would require considering the analogue of the Breitenlohner-Freedman (BF) bound for a warped ${\rm AdS}$ spacetime. We leave this to a future investigation.

\section*{Aknowledgements}
R.N. was supported by the European Union and the Czech Ministry of Education, Youth and
Sports (Project: MSCA Fellowship CZ FZU III - CZ.02.01.01/00/22$\_$010/0008598). MO is partially funded by Beca ANID de Doctorado grant 21222264.

\appendix

\section{Conventions}\label{appA}
We will denote the ten-dimensional curved indices by capital latin indices $M,N=0,\dots9$. Rigid indices will always be denoted with an underbar. 
For any $p$-form in 10 dimensions $\mathsf{F}_{p}$, we define
\begin{align}
|\mathsf{F}_{p}|^{2} & \equiv\frac{1}{p!}\mathsf{F}_{M_{1}\dots M_{p}}\mathsf{F}^{M_{1}\dots M_{p}}\,,\hspace{1.8cm}|\mathsf{F}_{p}|_{MN}^{2}\equiv\frac{1}{(p-1)!}\mathsf{F}_{MN_{1}\dots N_{p-1}}\mathsf{F}_{N}{}^{N_{1}\dots N_{p-1}}\,,\label{squareform}\\
\star\mathsf{F}_{p} & \equiv\frac{\sqrt{-g}}{p!(D-p)!}\dd x^{M_{1}\dots M_{D-p}}\epsilon_{M_{1}\dots M_{D-p}N_{1}\dots N_{p}}\mathsf{F}^{N_{1}\dots N_{p}}\,,
\end{align}
with $\epsilon_{12\dots D}=-\epsilon^{12\dots D}=1$. Moreover, we define $\dd x^{M_0\ldots M_9}=-\epsilon^{M_0\ldots M_9}\dd^{10}x$. The slash operation acts on
any $p$-form $\mathsf{F}_{p}$ as follows
\begin{align}\label{slashed}
\not{\mathsf{F}}_{p}=(\mathsf{F}_{p})_{/} & \equiv\frac{1}{p!}F_{M_{1}\dots M_{p}}\hat\Gamma^{M_{1}\dots M_{p}}\,.
\end{align}
The ten-dimensional Dirac matrices $\hat\Gamma^{\underline M}$ satisfy the Clifford algebra
\begin{align}
    \{\hat\Gamma^{\underline M},\hat\Gamma^{\underline M}\}=2\eta^{\underline{MN}}\mathds 1_{32}\,,
\end{align}
where $\eta^{\underline{MN}}$ is the flat metric in mostly plus signature. We will consider the following basis, adapted to the dimensional reduction, to be described in Appendix \ref{appB},
\begin{align}
\hat{\Gamma}^{\underline{\mu}} & =\gamma^{\underline{\mu}}\otimes\Gamma_{*4}\,,\hspace{1.5cm}\hat{\Gamma}^{\underline{a}+5}=\mathds{1}_{8\times8}\otimes\Gamma^{\underline{a}}\,.
\end{align}
Here the ten-dimensional index splits as $M=(\mu,a)$, with $\mu=0,\ldots,5$, $a=1,2,3,4$ and 
\begin{align}
\gamma^{0}&=\ri\sigma_{1}\otimes\mathds1_{4}\,,\qquad\gamma^{1}=-\sigma_{2}\otimes\sigma_{1}\otimes\mathds1_{2}\,,\qquad\gamma^{2}=-\sigma_{2}\otimes\sigma_{2}\otimes\mathds1_{2}\,,\nonumber\\
\gamma^{3}&=-\sigma_{2}\otimes\sigma_{3}\otimes\sigma_{1}\,,\qquad\gamma^{4}=-\sigma_{2}\otimes\sigma_{3}\otimes\sigma_{2}\,,\qquad\gamma^{5}=-\sigma_{2}\otimes\sigma_{3}\otimes\sigma_{3}\,, \nonumber\\
\Gamma^{1} & =\sigma_{1}\otimes\mathds1_{2}\,,\qquad\Gamma^{i}=-\sigma_{2}\otimes\sigma_{i-1}\,,\qquad i=2,3,4\,,
\end{align}
where $\sigma_{1,2,3}$ are the usual Pauli matrices. The above Dirac matrices close the Clifford algebra in six and four dimensions, with Lorentzian and Euclidean signature, respectively
\begin{align}
    \{\gamma^{\underline{\mu}},\gamma^{\underline\nu}\}=2\eta^{\underline{\mu\nu}}\mathds 1_{8}, \qquad \{\Gamma^{\underline{a}},\Gamma^{\underline b}\}=2\delta^{\underline{ab}}\mathds 1_{4}\,.
\end{align}
Relations analogous to \eqref{squareform} and \eqref{slashed} also hold for six-dimensional fields.\\
We define the 10-dimensional chirality and charge conjugation matrices as
\begin{align}
    \hat \Gamma_*&=-\gamma_{*6}\otimes\Gamma_{*4}=\hat\Gamma^0\hat\Gamma^1\hat\Gamma^2\hat\Gamma^3\hat\Gamma^4\hat\Gamma^5\hat\Gamma^6\hat\Gamma^7\hat\Gamma^8\hat\Gamma^9\,, \qquad
    \hat C=C_{6}\otimes C_{4}\,,
\end{align}
in terms of the six and four-dimensional quantities
\begin{align}
\gamma_{*6} & =-\gamma^{0}\gamma^{1}\gamma^{2}\gamma^{3}\gamma^{4}\gamma^{5}\,, \qquad \Gamma_{*4} =\Gamma^{1}\Gamma^{2}\Gamma^{3}\Gamma^{4}\,,\nonumber\\
C_{6} & =\ri \gamma^1\gamma^3\gamma^5\,, \qquad 
C_{4} =-\Gamma^{1}\Gamma^{3}\,.
\end{align}

\section{Details on the reduction}\label{appB}
In order to fix the notations, {which agree with those in \cite{Tomasiello:2022dwe}, we recall the basic facts about type IIB supergravity. The bosonic fields are the 10-dimensional metric (in the Einstein frame) $g_{MN}$, the dilaton scalar field $\Phi$, and the Kalb-Ramond 2-form $B_{2}$.  RR fields are even-form potentials $C_{0},C_{2},C_{4}$. 
The pseudo-action of type IIB supergravity reads:
\begin{align}\label{IIBaction}
I_{\mathrm{IIB}} & =\frac{1}{2\kappa^{2}}\int\dd^{10}x\sqrt{-g}(R-\frac{1}{2}\partial_{M}\Phi\partial^{M}\Phi-\frac{1}{2}e^{-\Phi}|H_{3}|^{2}-\frac{1}{2}e^{2\Phi}|\mathsf{F}_{1}|^{2}-\frac{1}{2}e^{\Phi}|\mathsf{F}_{3}|^{2}-\frac{1}{4}|\mathsf{F}_{5}|^{2})\nonumber \\
 & +\frac{1}{4\kappa^{2}}\int C_{4}\wedge H_{3}\wedge\mathsf{F}_{3}\,,
\end{align}
where 
\begin{align}\label{FDefs}
\mathsf{F}_{1} & =\dd C_{0}\,,\hspace{1.2cm}\mathsf{F}_{3}=\dd C_{2}-C_{0}\wedge H_{3}\,,\hspace{1.2cm}\mathsf{F}_{5}=\dd C_{4}-C_{2}\wedge H_{3}\,,
\end{align}
and $\mathsf{F}_{5}$ is required to be self-dual:
$$\mathsf{F}_{5}=*\mathsf{F}_{5}\,.$$
The equations of motion are
\begin{align}
0&=\nabla_M\partial^M\Phi+\frac{e^{-\Phi}}{2}\,|H_3|^2-\frac{e^{\Phi}}{2}\,|\mathsf{F}_3|^2-{e^{2\Phi}}\,|\mathsf{F}_1|^2\,, \nonumber\\
0&=\nabla_M\left(e^{2\Phi}\partial^MC_{0}\right)+\frac{e^\Phi}{3!}\,H_{3\,MNP}\,\mathsf{F}_3^{MNP}\,, \nonumber\\
0&=\dd\left[e^{\Phi}\,\star \mathsf{F}_3\right]+H_3\wedge \mathsf{F}_5\,,\nonumber\\
0&=\dd\left[e^{-\Phi}\,\star H_3-e^{\Phi}\,C_{0}\,\star \mathsf{F}_3\right]+\mathsf{F}_5\wedge (\mathsf{F}_3+C_{0}\,H_3)\,,\nonumber\\
0&=R_{MN} -\frac{1}{2}\partial_{M}\Phi\partial_{N}\Phi-\frac{e^{2\Phi}}{2}\,\partial_{M}C_{0}\partial_{N}C_{0}-\frac{e^{-\Phi}}{2}\left(|H_{3}|_{MN}^{2}-\frac{1}{4}\,g_{MN}\,|H_{3}|^2\right)\nonumber\\
&-\frac{e^{\Phi}}{2}\left(|\mathsf{F}_{3}|_{MN}^{2}-\frac{1}{4}\,g_{MN}\,|\mathsf{F}_{3}|^2\right)-\frac{1}{4}|\mathsf{F}_{5}|_{MN}^{2}\,.
\end{align}
The fermionic fields of type IIB supergravity are given by two gravitini $\boldsymbol{\Psi}_M=(\psi_M^{\alpha A})$ and two spin 1/2 fields $\boldsymbol{\lambda}=(\lambda^{\alpha A})$, with $\alpha=1,\ldots 32$ and $A=1,2$ a $\mathrm{SO}(2)$ doublet index. They are both Majorana-Weyl spinors, such that
\begin{align}
   \hat\Gamma_* \otimes\mathds 1_2\boldsymbol{\Psi}_M&=\boldsymbol{\Psi}_M\,,\qquad \bar{\boldsymbol{\Psi}}_M=\boldsymbol{\Psi}_M^T \hat C\otimes\mathds 1_2\,,\nonumber\\
    \hat\Gamma_*\otimes\mathds 1_2\boldsymbol{\lambda}&=-\boldsymbol{\lambda}\,,\qquad \bar{\boldsymbol{\lambda}}=\boldsymbol{\lambda}^T \hat C\otimes\mathds 1_2\,, \label{pseudo-majorana condition}
\end{align}
where $\mathds 1_2$ acts on the R-symmetry index, which is suppressed.
The supersymmetry variations of the spinors in the Einstein frame are
\begin{align}
\delta\boldsymbol{\lambda} & =\frac{1}{2}(\dd\Phi-e^{\Phi}\mathsf{F}_{1})_{/}\,\boldsymbol{\epsilon}-\frac{1}{4}(e^{-\frac{\Phi}{2}}H_{3}\sigma_{3}+e^{\frac{\Phi}{2}}\mathsf{F}_{3}\sigma_{1})_{/}\,\boldsymbol{\epsilon}\,,\\
\delta\boldsymbol{\Psi}_{M} & =\partial_M\boldsymbol{\epsilon}+\frac{1}{4}\omega_{M \underline{MN}}\Gamma^{\underline{MN}}\,\boldsymbol{\epsilon}+\frac{1}{4}e^{\Phi}\mathsf{F}_{M}\ri\sigma_{2}\,\boldsymbol{\epsilon}+\frac{1}{16}\not{\mathsf{F}}_{5}\ri\sigma_{2}\hat{\Gamma}_{M}\,\boldsymbol{\epsilon}\nonumber\\
& +\frac{1}{96}(e^{-\frac{\Phi}{2}}H_{NPQ}\sigma_{3}-e^{\frac{\Phi}{2}}\mathsf{F}_{NPQ}\sigma_{1})(\hat{\Gamma}_{M}\hat{\Gamma}^{NPQ}-12\delta_{M}^{N}\hat{\Gamma}^{PQ})\,\boldsymbol{\epsilon}\,,
\end{align}
where $\boldsymbol{\epsilon}$ is a Majorana-Weyl spinor with the same chirality as $\boldsymbol{\Psi}_M$.

Let us now briefly review the dimensional reduction of type IIB supergravity on an a $T^{4}/\bZ_{2}$-orientifold, i.e. in the presence of $O_{5}$-planes transverse to the internal 4-torus (and of corresponding D5 branes), restricting to the closed string zero-modes only. The coordinates $x^M$, parametrizing the ten-dimensional spacetime $\hM_{10}=\hM_{6}\times T^{4}/\bZ_{2}$, split as follows $x^{M}=(x^{\mu},y^{a})$, with $x^{\mu}$ and $y^{a}$ spanning the six-dimensional non-compact manifold and the 4-torus, respectively.
The $\bZ_{2}$ group is generated by an involution, $\sigma$, defined as the product of the Type IIB string world-sheet parity, $\Omega$, combined with a parity on the internal coordinates $I_{4}$:
\begin{align}
\sigma=\Omega I_{4}\,,\qquad\text{where}\qquad I_{4}:y^{a}\mapsto-y^{a}\,.
\end{align}
The ansatz for the metric is 
\begin{align}
\dd s^{2} & =\Delta^{2}g_{\mu\nu}\dd x^{\mu}\otimes\dd x^{\nu}+\upgamma_{ab}V^{a}\otimes V^{b}\,,\,\,\,V^{a}=\dd y^{a}+\mathscr{G}^{a}\,,
\end{align}
where $\mathscr{G}^{a}=\mathscr{G}^{a}_\mu(x^\nu)\,dx^\mu$ are the Kaluza-Klein vectors. To have the $D=6$ action written in the Einstein frame, we have to fix $\Delta^2=\upgamma^{-\frac{1}{4}}$, where $\upgamma\equiv {\rm det}(\upgamma_{ab})$. Just as for ordinary toroidal dimensional reductions, we expand all the fields in $dx^\mu,\,V^a$ and their duals, and assume the coefficients of the expansion only to depend on $x^\mu$. We now consider the consistent truncation of the theory to the invariant sector under the action of $\sigma$. Recalling that the parity properties of Type IIB fields under $\Omega$ are 
\begin{align}
\Omega-\mathrm{even}&:\qquad g_{MN}\,,\qquad\phi\,,\qquad C_{MN}\,,\nonumber\\
\Omega-\mathrm{odd}&: \qquad B_{MN}\,,\qquad C_{0}\,,\qquad C_{MNPQ}\,, 
\end{align}
the fields after the truncation are 
\begin{align}\label{FieldTrunc}
\mathscr{G}^{a} &=0\,,\,\,\Rightarrow\,\,\,V^a=\dd y^a\,,\nonumber\\
C_2 & =C_{(2,0)}+C_{(0,2)}\,,\,\,C_{(2,0)}=C_{(2)}\equiv \frac{1}{2}C_{\mu\nu}\dd x^{\mu\nu}\,,\,\,C_{(0,2)}\equiv\frac{1}{2}C_{ab}\dd y^{ab}\,,\nonumber\\
B_2 & =B_{(1,1)}\equiv B_{(1)a}\,\dd y^a\,,\nonumber\\
C_4 & =C_{(3,1)}+C_{(1,3)}\,,\,\,C_{(3,1)}\equiv \frac{1}{3!}C_{\mu\nu\rho a}\wedge dx^{\mu\nu\rho}\wedge \dd y^{a}\,,\,\,\,C_{(1,3)}\equiv\frac{1}{3!}\,C_{(1)abc}\wedge \dd y^{abc} \,,\nonumber\\
C_{0} &= 0\,,
\end{align}
where we have used the short-hand notation $\dd y^{a_{1}\dots a_{p}}\equiv \dd y^{a_{1}}\wedge\dots\wedge \dd y^{a_{p}}$ and $\dd x^{\mu_{1}\dots\mu_{n}} \equiv \dd x^{\mu_{1}}\wedge\dots \wedge \dd x^{\mu_{n}}$. In the above ansatz, $C_{(2)}$ denotes the $D=6$ rank-2 tensor field, while $B_{(1)a},\,C_{(1)abc}$ will be identified with the eight vector fields of the six-dimensional theory. \par
We perform the dimensional reduction on the $T^4/\mathbb{Z}_2$-orientifold by following the procedure described in  \cite{Berg:2003ri}. In this case, we plug the above ansatz for the bosonic fields in the pseudo-action and add a Lagrangian multiplier in order to dualize $C_{(3,1)}$ to $\hat{C}_{(1,3)}$. Then, using the self-duality of $\mathsf{F}_{(5)}$, we identify this dual field with the component ${C}_{(1,3)}$. Using \eqref{FDefs}, and \eqref{FieldTrunc}, the self-duality equation $\star\mathsf{F}_{5} = \mathsf{F}_{5}$ reduces to
\begin{align}\label{selfduality}
\star\left(\dd C_{(4,1)}-C_{(2,0)}\wedge\dd B_{(2,1)}\right) & =\dd C_{(2,3)}-C_{(0,2)}\wedge\dd B_{(2,1)}\,.
\end{align}
In order to dualize the $C_{(3,1)}$ component for the 4-form, we replace everywhere in the pseudo-action $\dd C_{(3,1)}$ by an auxiliary field $\mathring{\mathsf{F}}^{4,1}$. We denote by $I'_{{\rm IIB}}$ the resulting action. Then, we sum a Lagrange multiplier term, so to obtain
\begin{equation}
    I'=I'_{{\rm IIB}}-\frac{1}{4\kappa^2}\,\int \mathring{\mathsf{F}}_{(4,1)}\wedge \dd\hat{C}_{(1,3)}\,.
\end{equation}
Varying with respect to the Lagrange multiplier $\hat{C}_{(1,3)}$ we find:
$$\frac{\delta I'}{\delta \hat{C}_{(1,3)}}=0\,\,\Rightarrow\,\,\,\dd \mathring{\mathsf{F}}_{(4,1)}=0\,\,\Rightarrow\,\,\,\mathring{\mathsf{F}}_{(4,1)}=\dd C_{(3,1)}\,.$$
In other words, $\hat{C}_{(1,3)}$ is the Lagrange multiplier implementing the Bianchi identity for $\mathring{\mathsf{F}}_{(4,1)}$.
Extremizing $I'$ with respect to $\mathring{\mathsf{F}}_{(4,1)}$, on the other hand, yields to following equation:
\begin{align}\label{selfduality2}
\frac{\delta I'}{\delta \mathring{\mathsf{F}}_{(4,1)}}=0\,\,\Rightarrow\,\,\,\star\left(\dd C_{(4,1)}-C_{(2,0)}\wedge\dd B_{(2,1)}\right) & =\dd \hat{C}_{(2,3)}-C_{(0,2)}\wedge\dd B_{(2,1)}\,,
\end{align}
which, in light of \eqref{selfduality}, allows us to identify $\hat{C}_{(1,3)}={C}_{(1,3)}$. Solving this equation with respect to $\mathring{\mathsf{F}}_{(4,1)}$ and replacing into $I'$, the terms coming from $\star \mathsf{F}_5\wedge \mathsf{F}_5$ vanish.
The resulting action $I$ reads:
\begin{align}
    I=&\frac{1}{2\kappa^2}\,\int\left[R\star 1-\frac{1}{2}\star\dd \Phi\wedge \dd \Phi-\frac{e^{\Phi}}{2}\star\mathsf{F}_3\wedge \mathsf{F}_3-\frac{e^{-\Phi}}{2}\star H_3\wedge H_3+\right.\nonumber\\&\left.-\frac{1}{2}\star\left(\dd{C}_{(2,3)}-C_{(0,2)}\wedge\dd B_{(2,1)}\right)\wedge \left(\dd {C}_{(2,3)}-C_{(0,2)}\wedge\dd B_{(2,1)}\right)+C_{(2,0)}\wedge \dd C_{(2,3)}\wedge \dd B_{(2,1)}\right]
\end{align}
To obtain the $D=6$ action, we still have to express the Ricci scalar $R$ of the $D=10$ metric in terms of that associated with the $D=6$ metric $g_{\mu\nu}$ in the Einstein frame and denoted by $\mathcal{R}$. We also use the fact that the fields do not depend on the torus coordinates in order to factor out the volume of $T^4$. Defining then $\kappa_6^2\equiv \kappa^2/\int_{T^4}d^4y$ and $H_{(2)a}=\dd B_{(2)a}$, we find  
\begin{align}
I_{\mathrm{6D}} & =\frac{1}{2\kappa_{6}^{2}}\int[\mathcal{R}\star1-\frac{1}{2}e^{\phi}\star F_{(3)}\wedge F_{(3)}+ \nonumber\\
& -\frac{1}{2}e^{-\frac{1}{2}\phi}(\hat{\gamma}^{ab}\star H_{(2)a}\wedge H_{(2)b}+\hat{\gamma}_{ab}\star\check{\mathcal{F}}_{(2)}^{a}\wedge\check{\mathcal{F}}_{(2)}^{b})+C_{(2)}\wedge\dd\check{\mathcal{C}}_{(2)}^{a}\wedge H_{(2)a}+\nonumber \\
& -\frac{1}{4}\star\dd\phi\wedge\dd\phi+\frac{1}{4}\star\dd\hat{\gamma}_{ab}\wedge\dd\hat{\gamma}^{ab}-\frac{1}{4}\hat{\gamma}\hat{\gamma}^{ac}\hat{\gamma}^{bd}\star\dd C_{(1)ab}\wedge\dd C_{(1)cd}]\,,
\end{align}
where the Hodge duality is computed with the six-dimensional metric $g_{\mu\nu}$ and we have applied the following field redefinitions:
\begin{align}
\upgamma_{ab} &\equiv e^{\phi/4}\hat{\gamma}^{-3/8}\hat{\gamma}_{ab}\,,\hspace{2.6cm}\Phi \equiv\frac{\phi}{2} + \frac{1}{4}\log(\hat{\gamma})\,,\hspace{1.8cm} \label{B.20}\\
\check{\mathcal{F}}^{d}_{(2)}  &\equiv\dd\check{\mathcal{C}}^{d}_{(2)}-\frac{1}{2}\,H_{(2)a}\,C_{bc}\varepsilon^{abcd}\,,\hspace{1.2cm} \check{\mathcal{C}}_{(1)}^{d} \equiv \frac{1}{3!}\epsilon^{abcd}C_{(1)abc} \\
\Delta& =e^{-\phi/8} \hat \gamma ^{1/16}\,,\label{defDelta}
\end{align}
reproducing finally \eqref{bosonicaction}.

Here we summarize the ten-dimensional fields in Einstein frame in terms of the six-dimensional ones
\begin{align}\label{uplift type IIB}
    \dd s^2_{10} & = e^{-\phi/4}\hat\gamma^{1/8} \dd s^2_6 + e^{\phi/4}\hat \gamma^{-3/8} \hat \gamma_{ab} \dd y^a \dd y^b\,, \nonumber \\
    \mathsf{F}_5 & = G_5 + \star_{10} G_5 \, , \qquad G_5 \equiv \dd C_{(2,3)}-C_{(0,2)}\wedge\dd B_{(2,1)}= \frac{1}{3!}  \check{\mathcal{F}}_{(2)}^d \epsilon_{abcd} \dd y^{a} \wedge \dd y^{b} \wedge  \dd y^{c} \,,\nonumber \\
    \mathsf{F}_3 & = \dd C_2 + \frac12 \dd C_{a b} \wedge \dd y^a \wedge  \dd y^b \,,\,\,\,
    \mathsf{F}_1  = 0\,, \nonumber \\
    \Phi & = \frac{\phi}{2} + \frac14 \log \hat \gamma \, , \nonumber\\
    H_3 & = H_{(2)\,a} \wedge \dd y ^a \, .
\end{align}

\bibliographystyle{utphys}
\bibliography{refs.bib}

\providecommand{\href}[2]{#2}\begingroup\raggedright\begin{thebibliography}{10}

\bibitem{Maldacena:1998bw}
J.~M. Maldacena and A.~Strominger, ``{A}d{S}$_3$ black holes and a stringy exclusion principle,'' \href{http://dx.doi.org/10.1088/1126-6708/1998/12/005}{{\em JHEP} {\bfseries 12} (1998) 005},
\href{http://arxiv.org/abs/hep-th/9804085}{{\ttfamily arXiv:hep-th/9804085 [hep-th]}}.

\bibitem{Deger:1998nm}
S.~Deger, A.~Kaya, E.~Sezgin, and P.~Sundell, ``Spectrum of ${D} = 6$, ${N}=4b$ supergravity on {AdS}$_3\times {S}^3$,'' \href{http://dx.doi.org/10.1016/S0550-3213(98)00555-0}{{\em Nucl.Phys.} {\bfseries B536} (1998) 110--140},
\href{http://arxiv.org/abs/hep-th/9804166}{{\ttfamily arXiv:hep-th/9804166 [hep-th]}}.

\bibitem{deBoer:1998kjm}
J.~de~Boer, ``Six-dimensional supergravity on ${S}^3 \times {AdS}_3$ and $2d$ conformal field theory,'' \href{http://dx.doi.org/10.1016/S0550-3213(99)00160-1}{{\em Nucl. Phys.} {\bfseries B548} (1999) 139--166},
\href{http://arxiv.org/abs/hep-th/9806104}{{\ttfamily arXiv:hep-th/9806104 [hep-th]}}.

\bibitem{Eberhardt:2019ywk}
L.~Eberhardt, M.~R. Gaberdiel, and R.~Gopakumar, ``Deriving the {AdS$_{3}$/CFT$_{2}$} correspondence,'' \href{http://dx.doi.org/10.1007/JHEP02(2020)136}{{\em JHEP} {\bfseries 02} (2020) 136}, \href{http://arxiv.org/abs/1911.00378}{{\ttfamily arXiv:1911.00378 [hep-th]}}.

\bibitem{Anninos:2008fx}
D.~Anninos, W.~Li, M.~Padi, W.~Song, and A.~Strominger, ``Warped {AdS}$_3$ black holes,'' \href{http://dx.doi.org/10.1088/1126-6708/2009/03/130}{{\em JHEP} {\bfseries 0903} (2009) 130},
\href{http://arxiv.org/abs/0807.3040}{{\ttfamily arXiv:0807.3040 [hep-th]}}.

\bibitem{Guica:2010sw}
M.~Guica, K.~Skenderis, M.~Taylor, and B.~C. van Rees, ``Holography for {S}chr{\"o}dinger backgrounds,'' \href{http://dx.doi.org/10.1007/JHEP02(2011)056}{{\em JHEP} {\bfseries 02} (2011) 056}, \href{http://arxiv.org/abs/1008.1991}{{\ttfamily arXiv:1008.1991 [hep-th]}}.

\bibitem{Song:2011sr}
W.~Song and A.~Strominger, ``Warped {AdS}$_3/$dipole-{CFT} duality,'' \href{http://dx.doi.org/10.1007/JHEP05(2012)120}{{\em JHEP} {\bfseries 05} (2012) 120}, \href{http://arxiv.org/abs/1109.0544}{{\ttfamily arXiv:1109.0544 [hep-th]}}.

\bibitem{Detournay:2012pc}
S.~Detournay, T.~Hartman, and D.~M. Hofman, ``Warped conformal field theory,'' \href{http://dx.doi.org/10.1103/PhysRevD.86.124018}{{\em Phys. Rev. D} {\bfseries 86} (2012) 124018}, \href{http://arxiv.org/abs/1210.0539}{{\ttfamily arXiv:1210.0539 [hep-th]}}.

\bibitem{Afshar:2015wjm}
H.~Afshar, S.~Detournay, D.~Grumiller, and B.~Oblak, ``Near-horizon geometry and warped conformal symmetry,'' \href{http://dx.doi.org/10.1007/JHEP03(2016)187}{{\em JHEP} {\bfseries 03} (2016) 187}, \href{http://arxiv.org/abs/1512.08233}{{\ttfamily arXiv:1512.08233 [hep-th]}}.

\bibitem{Israel:2004vv}
D.~Israel, C.~Kounnas, D.~Orlando, and P.~M. Petropoulos, ``Electric/magnetic deformations of {$S^3$} and {AdS$_3$}, and geometric cosets,'' \href{http://dx.doi.org/10.1002/prop.200410190}{{\em Fortsch. Phys.} {\bfseries 53} (2005) 73--104}, \href{http://arxiv.org/abs/hep-th/0405213}{{\ttfamily arXiv:hep-th/0405213}}.

\bibitem{Detournay:2005fz}
S.~Detournay, D.~Orlando, P.~M. Petropoulos, and P.~Spindel, ``Three-dimensional black holes from deformed anti-de {S}itter,'' \href{http://dx.doi.org/10.1088/1126-6708/2005/07/072}{{\em JHEP} {\bfseries 07} (2005) 072}, \href{http://arxiv.org/abs/hep-th/0504231}{{\ttfamily arXiv:hep-th/0504231}}.

\bibitem{Azeyanagi:2012zd}
T.~Azeyanagi, D.~M. Hofman, W.~Song, and A.~Strominger, ``The spectrum of strings on warped {AdS}$_3 \times$ {S}$^3$,'' \href{http://dx.doi.org/10.1007/JHEP04(2013)078}{{\em JHEP} {\bfseries 04} (2013) 078}, \href{http://arxiv.org/abs/1207.5050}{{\ttfamily arXiv:1207.5050 [hep-th]}}.

\bibitem{Bobev:2011qx}
N.~Bobev and B.~C. van Rees, ``Schr{\"o}dinger deformations of {$AdS_3 \times S^3$},'' \href{http://dx.doi.org/10.1007/JHEP08(2011)062}{{\em JHEP} {\bfseries 08} (2011) 062}, \href{http://arxiv.org/abs/1102.2877}{{\ttfamily arXiv:1102.2877 [hep-th]}}.

\bibitem{El-Showk:2011euy}
S.~El-Showk and M.~Guica, ``{Kerr/CFT}, dipole theories and nonrelativistic {CFTs},'' \href{http://dx.doi.org/10.1007/JHEP12(2012)009}{{\em JHEP} {\bfseries 12} (2012) 009}, \href{http://arxiv.org/abs/1108.6091}{{\ttfamily arXiv:1108.6091 [hep-th]}}.

\bibitem{Eloy:2023acy}
C.~Eloy and G.~Larios, ``{Nonsupersymmetric stable marginal deformations in AdS$_3$/CFT$_2$},'' \href{http://dx.doi.org/10.1103/PhysRevD.108.L121901}{{\em Phys. Rev. D} {\bfseries 108} no.~12, (2023) L121901}, \href{http://arxiv.org/abs/2309.03261}{{\ttfamily arXiv:2309.03261 [hep-th]}}.

\bibitem{Eloy:2024lwn}
C.~Eloy and G.~Larios, ``Charting the conformal manifold of holographic {CFT$_2$'s},'' \href{http://dx.doi.org/10.21468/SciPostPhys.17.4.123}{{\em SciPost Phys.} {\bfseries 17} no.~4, (2024) 123}, \href{http://arxiv.org/abs/2405.17542}{{\ttfamily arXiv:2405.17542 [hep-th]}}.

\bibitem{Deger:2024xnd}
N.~S. Deger, C.~A. Deral, A.~Saha, and O.~Sar\i{}o\u{g}lu, ``Rotating {AdS}$_3\times$ {S$^{3}$} and dyonic strings from 3 dimensions,'' \href{http://dx.doi.org/10.1007/JHEP10(2024)185}{{\em JHEP} {\bfseries 10} (2024) 185}, \href{http://arxiv.org/abs/2408.03197}{{\ttfamily arXiv:2408.03197 [hep-th]}}.

\bibitem{Deger:2024obg}
N.~S. Deger and C.~A. Deral, ``Timelike supersymmetric solutions of {$D=3, N=4$} supergravity,'' \href{http://dx.doi.org/10.1103/PhysRevD.111.046007}{{\em Phys. Rev. D} {\bfseries 111} no.~4, (2025) 046007}, \href{http://arxiv.org/abs/2411.04437}{{\ttfamily arXiv:2411.04437 [hep-th]}}.

\bibitem{Angelantonj:2002ct}
C.~Angelantonj and A.~Sagnotti, ``{Open strings},'' \href{http://dx.doi.org/10.1016/S0370-1573(02)00273-9}{{\em Phys. Rept.} {\bfseries 371} (2002) 1--150}, \href{http://arxiv.org/abs/hep-th/0204089}{{\ttfamily arXiv:hep-th/0204089}}. [Erratum: Phys.Rept. 376, 407 (2003)].

\bibitem{Eloy:2021fhc}
C.~Eloy, G.~Larios, and H.~Samtleben, ``Triality and the consistent reductions on {AdS$_{3}\times S^{3}$},'' \href{http://dx.doi.org/10.1007/JHEP01(2022)055}{{\em JHEP} {\bfseries 01} (2022) 055}, \href{http://arxiv.org/abs/2111.01167}{{\ttfamily arXiv:2111.01167 [hep-th]}}.

\bibitem{Hoare:2022asa}
B.~Hoare, F.~K. Seibold, and A.~A. Tseytlin, ``{Integrable supersymmetric deformations of AdS$_{3}$\texttimes{} S$^{3}$\texttimes{} T$^{4}$},'' \href{http://dx.doi.org/10.1007/JHEP09(2022)018}{{\em JHEP} {\bfseries 09} (2022) 018}, \href{http://arxiv.org/abs/2206.12347}{{\ttfamily arXiv:2206.12347 [hep-th]}}.

\bibitem{Chow:2009km}
D.~D.~K. Chow, C.~N. Pope, and E.~Sezgin, ``{Classification of solutions in topologically massive gravity},'' \href{http://dx.doi.org/10.1088/0264-9381/27/10/105001}{{\em Class. Quant. Grav.} {\bfseries 27} (2010) 105001}, \href{http://arxiv.org/abs/0906.3559}{{\ttfamily arXiv:0906.3559 [hep-th]}}.

\bibitem{Deger:2013yla}
N.~S. Deger, A.~Kaya, H.~Samtleben, and E.~Sezgin, ``Supersymmetric warped {AdS} in extended topologically massive supergravity,'' \href{http://dx.doi.org/10.1016/j.nuclphysb.2014.04.011}{{\em Nucl. Phys. B} {\bfseries 884} (2014) 106--124}, \href{http://arxiv.org/abs/1311.4583}{{\ttfamily arXiv:1311.4583 [hep-th]}}.

\bibitem{Bieliavsky:2024hus}
P.~Bieliavsky, P.~Spindel, and R.~Wutte, ``Aspects of warped {AdS$_3$} geometries,'' \href{http://arxiv.org/abs/2410.09688}{{\ttfamily arXiv:2410.09688 [gr-qc]}}.

\bibitem{Clement:2009gq}
G.~Clement, ``Warped {AdS$_3$} black holes in new massive gravity,'' \href{http://dx.doi.org/10.1088/0264-9381/26/10/105015}{{\em Class. Quant. Grav.} {\bfseries 26} (2009) 105015}, \href{http://arxiv.org/abs/0902.4634}{{\ttfamily arXiv:0902.4634 [hep-th]}}.

\bibitem{Corral:2024xfv}
C.~Corral, D.~Flores-Alfonso, G.~Giribet, and J.~Oliva, ``Self-gravitating solutions in {Y}ang-{M}ills-{C}hern-{S}imons theory coupled to {3D} massive gravity,'' \href{http://dx.doi.org/10.1140/epjc/s10052-024-13326-z}{{\em Eur. Phys. J. C} {\bfseries 84} no.~9, (2024) 959}, \href{http://arxiv.org/abs/2404.15569}{{\ttfamily arXiv:2404.15569 [hep-th]}}.

\bibitem{Setare:2017xlu}
M.~R. Setare and H.~Adami, ``{Near Horizon Geometry of Warped Black Holes in Generalized Minimal Massive Gravity},'' \href{http://arxiv.org/abs/1711.08344}{{\ttfamily arXiv:1711.08344 [hep-th]}}.

\bibitem{Nam:2018gju}
S.~Nam and J.-D. Park, ``{Warped AdS$_3$ black hole in minimal massive gravity with first order formalism},'' \href{http://dx.doi.org/10.1103/PhysRevD.98.124034}{{\em Phys. Rev. D} {\bfseries 98} no.~12, (2018) 124034}, \href{http://arxiv.org/abs/1808.00744}{{\ttfamily arXiv:1808.00744 [hep-th]}}.

\bibitem{Deger:2024dbz}
N.~S. Deger, J.~Rosseel, and H.~Samtleben, ``{The general supersymmetric solution of minimal massive supergravity},'' \href{http://dx.doi.org/10.1016/j.physletb.2024.139183}{{\em Phys. Lett. B} {\bfseries 860} (2025) 139183}, \href{http://arxiv.org/abs/2410.07964}{{\ttfamily arXiv:2410.07964 [hep-th]}}.

\bibitem{Andrianopoli:2023dfm}
L.~Andrianopoli, B.~L. Cerchiai, R.~Noris, L.~Ravera, M.~Trigiante, and J.~Zanelli, ``New torsional deformations of locally {AdS$_3$} space,'' \href{http://dx.doi.org/10.1103/PhysRevD.108.044011}{{\em Phys. Rev. D} {\bfseries 108} no.~4, (2023) 044011}, \href{http://arxiv.org/abs/2305.17168}{{\ttfamily arXiv:2305.17168 [hep-th]}}.

\bibitem{Andrianopoli:2024twc}
L.~Andrianopoli, R.~Noris, M.~Trigiante, and J.~Zanelli, ``Supersymmetric states in {Anti-de Sitter} {D=3} supergravity with chiral torsion,'' \href{http://dx.doi.org/10.1103/PhysRevLett.133.031602}{{\em Phys. Rev. Lett.} {\bfseries 133} no.~3, (2024) 031602}, \href{http://arxiv.org/abs/2404.12427}{{\ttfamily arXiv:2404.12427 [hep-th]}}.

\bibitem{Okumura}
M.~Okumura, ``{Some remarks on space with a certain contact structure},'' {\em Tohoku Math. J.} {\bfseries 14} (1962) 135.

\bibitem{Boyer:2004eh}
C.~P. Boyer, K.~Galicki, and P.~Matzeu, ``{On eta-Einstein Sasakian geometry},'' \href{http://dx.doi.org/10.1007/s00220-005-1459-6}{{\em Commun. Math. Phys.} {\bfseries 262} (2006) 177--208}, \href{http://arxiv.org/abs/math/0406627}{{\ttfamily arXiv:math/0406627}}.

\bibitem{Townsend:1983xs}
P.~K. Townsend, K.~Pilch, and P.~van Nieuwenhuizen, ``Selfduality in odd dimensions,''
{\em Phys. Lett.} {\bfseries 136B} (1984) 38.

\bibitem{Martelli:2013aqa}
D.~Martelli and A.~Passias, ``{The gravity dual of supersymmetric gauge theories on a two-parameter deformed three-sphere},'' \href{http://dx.doi.org/10.1016/j.nuclphysb.2013.09.012}{{\em Nucl. Phys. B} {\bfseries 877} (2013) 51--72}, \href{http://arxiv.org/abs/1306.3893}{{\ttfamily arXiv:1306.3893 [hep-th]}}.

\bibitem{Bobev:2016sap}
N.~Bobev, T.~Hertog, and Y.~Vreys, ``{The NUTs and Bolts of Squashed Holography},'' \href{http://dx.doi.org/10.1007/JHEP11(2016)140}{{\em JHEP} {\bfseries 11} (2016) 140}, \href{http://arxiv.org/abs/1610.01497}{{\ttfamily arXiv:1610.01497 [hep-th]}}.

\bibitem{Bueno:2018yzo}
P.~Bueno, P.~A. Cano, R.~A. Hennigar, and R.~B. Mann, ``{Universality of Squashed-Sphere Partition Functions},'' \href{http://dx.doi.org/10.1103/PhysRevLett.122.071602}{{\em Phys. Rev. Lett.} {\bfseries 122} no.~7, (2019) 071602}, \href{http://arxiv.org/abs/1808.02052}{{\ttfamily arXiv:1808.02052 [hep-th]}}.

\bibitem{Canfora:2023bug}
F.~Canfora and C.~Corral, ``{Gravitating anisotropic merons and squashed spheres in the three-dimensional Einstein-Yang-Mills-Chern-Simons theory},'' \href{http://dx.doi.org/10.1007/JHEP11(2023)146}{{\em JHEP} {\bfseries 11} (2023) 146}, \href{http://arxiv.org/abs/2309.15693}{{\ttfamily arXiv:2309.15693 [hep-th]}}.

\bibitem{Kraus:2011pf}
P.~Kraus and E.~Perlmutter, ``Universality and exactness of {S}chr\"odinger geometries in string and {M}-theory,'' \href{http://dx.doi.org/10.1007/JHEP05(2011)045}{{\em JHEP} {\bfseries 05} (2011) 045}, \href{http://arxiv.org/abs/1102.1727}{{\ttfamily arXiv:1102.1727 [hep-th]}}.

\bibitem{Astesiano:2024gzy}
D.~Astesiano, D.~Ruggeri, and M.~Trigiante, ``{U folds from geodesics in moduli space},'' \href{http://dx.doi.org/10.1103/PhysRevD.109.086018}{{\em Phys. Rev. D} {\bfseries 109} no.~8, (2024) 086018}, \href{http://arxiv.org/abs/2401.04209}{{\ttfamily arXiv:2401.04209 [hep-th]}}.

\bibitem{Moussa:2008sj}
K.~A. Moussa, G.~Clement, H.~Guennoune, and C.~Leygnac, ``Three-dimensional {C}hern-{S}imons black holes,'' \href{http://dx.doi.org/10.1103/PhysRevD.78.064065}{{\em Phys.Rev.} {\bfseries D78} (2008) 064065},
\href{http://arxiv.org/abs/0807.4241}{{\ttfamily arXiv:0807.4241 [gr-qc]}}.

\bibitem{Banados:1992gq}
M.~Banados, M.~Henneaux, C.~Teitelboim, and J.~Zanelli, ``{Geometry of the (2+1) black hole},'' \href{http://dx.doi.org/10.1103/PhysRevD.48.1506}{{\em Phys. Rev. D} {\bfseries 48} (1993) 1506--1525}, \href{http://arxiv.org/abs/gr-qc/9302012}{{\ttfamily arXiv:gr-qc/9302012}}. [Erratum: Phys.Rev.D 88, 069902 (2013)].

\bibitem{Tomasiello:2022dwe}
A.~Tomasiello, \href{http://dx.doi.org/10.1017/9781108635745}{{\em {Geometry of String Theory Compactifications}}}.
\newblock Cambridge University Press, 2022.

\bibitem{Berg:2003ri}
M.~Berg, M.~Haack, and B.~Kors, ``An orientifold with fluxes and branes via {T}-duality,'' {\em Nucl. Phys.} {\bfseries B669} (2003) 3--56,
\href{http://arxiv.org/abs/hep-th/0305183}{{\ttfamily hep-th/0305183}}.

\end{thebibliography}\endgroup

\end{document}